\documentclass[aps,prb,twocolumn,10pt,superscriptaddress,showpacs]{revtex4-2}
\usepackage{hyperref}
\usepackage{epsfig}
\usepackage{graphicx}
\usepackage{subfigure}
\usepackage{latexsym}
\usepackage{color}
\usepackage{fullpage}
\usepackage{dcolumn}
\usepackage{bm}
\usepackage[normalem]{ulem}
\usepackage{units}
\usepackage{amsmath}
\usepackage{booktabs}
\usepackage{multirow}
\usepackage{lineno}

\begin{document}
\title{Water, vacancies, and competing exchange interactions in Prussian blue analogues: a neutron diffraction study of field- and dehydration-driven magnetic transitions}

\author{N. S. Dhami}
\email{naveen.dhami@universite-paris-saclay.fr}
\affiliation{Université Paris-Saclay, CNRS, Laboratoire de Physique des Solides, UMR-8502, 91405, Orsay, France}
\affiliation{Synchrotron SOLEIL, L’Orme des Merisiers, Saint Aubin BP 48, 91192 Gif-sur-Yvette, France}

\author{C. V. Colin}
\affiliation{Institut Néel, Université Grenoble Alpes \& CNRS, Grenoble, 38042, France}

\author{V. Nassif}
\affiliation{Institut Néel, Université Grenoble Alpes \& CNRS, Grenoble, 38042, France}

\author{O. Fabelo}
\affiliation{Institut Laue-Langevin, 38000 Grenoble, France}

\author{T. Nait}
\affiliation{ICMMO, Université Paris Saclay, CNRS, 15 rue Georges Clémenceau, 91405 Orsay, France}

\author{A. Bleuzen}
\affiliation{ICMMO, Université Paris Saclay, CNRS, 15 rue Georges Clémenceau, 91405 Orsay, France}

\author {A. Bordage}
\email{amelie.bordage@universite-paris-saclay.fr}
\affiliation{ICMMO, Université Paris Saclay, CNRS, 15 rue Georges Clémenceau, 91405 Orsay, France}

\author{V. Balédent}
\email{victor.baledent@universite-paris-saclay.fr}
\affiliation{Université Paris-Saclay, CNRS, Laboratoire de Physique des Solides, UMR-8502, 91405, Orsay, France}
\affiliation{Institut universitaire de France (IUF)}

\date{\today}

\begin{abstract}
  We report a neutron diffraction study of the structural and magnetic properties of a family of ferro- and ferrimagnetic Prussian blue analogues (PBAs), $A_4$[Fe(CN)$_6$]$_{2.7}$ ($A$ = Co, Mn, Ni), Rb$_2$Ni$_4$[Fe(CN)$_6$]$_{3.3}$, and Mn$_4$[Cr(CN)$_6$]$_{2.7}$, as a function of temperature (2--450~K) and applied magnetic field. All bimetallic compounds of the Fm$\overline{3}$m family exhibit a broad diffuse feature at low scattering angle, which we identify, through comparison with the cation-stabilized Rb$_2$NiFe framework, as an intrinsic signature of correlated vacancies and their associated interstitial water. High-temperature diffraction reveals a continuous crossover from positive to negative thermal expansion in CoFe and MnFe upon dehydration, while NiFe remains structurally robust up to 450~K. At low temperature, all compounds order in a collinear ferrimagnetic state with propagation vector $\mathbf{k}=(0,0,0)$, except MnFe, which adopts a partially frustrated magnetic structure with $\mathbf{k}=(1,0,0)$. A moderate magnetic field of $B_c = 1.3$~T drives a spin reorientation in MnFe toward the collinear $\mathbf{k}=(0,0,0)$ ferrimagnetic state common to the other compounds; the same transition is independently induced by dehydration. A minimal Heisenberg model shows that this transition results from a near-compensation between antiferromagnetic Mn--Fe coupling and a geometrically frustrated antiferromagnetic Mn--Mn interaction on the face-centered-cubic Mn sublattice, placing MnFe in the vicinity of a magnetic compensation point. These results resolve a longstanding ambiguity in the interpretation of the Fe $K$-edge XMCD response of MnFe-based PBAs, and establish water content as a key parameter controlling both the structural and magnetic stability of this family of materials, with direct relevance to their use as battery electrodes.
\end{abstract}

\maketitle

\section{Introduction}
Prussian blue analogues (PBAs) are a remarkably versatile family of metal–cyanide framework materials whose importance comes from the combination of simple chemistry and extraordinary functional properties. Their open, cubic lattice can host metal ions, based on alkali cations, and water molecules, giving them tunable compositions, rich electronic/magnetic behaviors, and highly accessible porosity. Because of this chemical flexibility, PBAs can exhibit various properties from long-range magnetism to photo-switching, ion-storage for next-generation batteries \cite{yi2021structure, peng2022prussian}, selective ion trapping, and even multiscale porosity when grown as hollow or nano-structured particles or nanotubes \cite{song2022confinement}. This coexistence of structural simplicity with complex, emergent properties makes PBAs scientifically fascinating and technologically promising across catalysis \cite{xu2022pba}, energy storage \cite{song2021prussian}, hydrogen storage \cite{kaye2005hydrogen},  sensing \cite{jiang2021recent}, photomagnetism \cite{yao2024achievements, fornasieri2018magnetism}, and environmental remediation as waste water treatment \cite{song2022confinement, su2025ni}.

Even though PBAs were among the compounds to be extensively investigated, important details about their magnetic structure and the role that water plays in both the crystal lattice and magnetic behavior remain insufficiently explored. The effect of water removal on the crystal and magnetic properties of PBA compounds is of particular importance for battery-related research \cite{wang2022effect}, but it also manifests strikingly in their thermal expansion behavior: depending on their guest (water and cation) content, PBA frameworks can display either negative thermal expansion (NTE) \cite{adak2011thermal} or near-zero thermal expansion (ZTE) \cite{margadonna2004zero}, two properties of direct technological relevance since they allow materials to withstand heat dissipation and thermal shock without damaging the framework. This sensitivity of both the structural and magnetic properties to water content illustrates that a deeper, fundamental understanding of the interactions between water and the PBA framework is much needed to rationally improve these materials for their targeted applications.

PBAs are molecular compounds of general formula Y$_x$A$_4$[B(CN)$_6$]$_{(8+x)/3}$.nH$_2$O (Y=alkali cation; x=0-4; A,B= 3d TM), widely investigated as functional materials. However, they also constitute a large family of model compounds particularly interesting for spectroscopic developments, because whatever the TM, (i) they are isostructural (FCC structure, Fm-3m space group), and (ii) they are formed of A—NC—B linkages in the three direction of space and with a perfect alternating of the TM. Depending on the x value, the coordination sphere of the A TM may be completed by water molecules; these vacancies in the network are randomly distributed. Additionally, PBA present a dominant exchange interaction between the first metallic neighbors. Using a series of model PBAs, Bleuzen et al. worked on disentangling the physical effects underlying the TM $K$-edge XMCD signals. Using bimetallic
A$_4$[B(CN)$_6]_{2.7}$ and trimetallic Rb$_2$(Co$_{1-x}$Ni$_x$)$_4$[Fe(CN)$_6$]$_{3.3}$ PBAs, unprecedented insights were gained through a purely experimental approach combining macroscopic and local characterization TM-$K$ edge x-ray absorption spectroscopy (XAS) and x-ray magnetic circular dicrhoism (XMCD) \cite{n2022toward, n2022interplay, n2024interplay}. Here we present a complete magnetic structure determination with neutron diffraction, necessary to generalize and understand the relations of XMCD with magnetism in these compounds. This study also highlights, why MnFe Fe $K$-edge XMCD results were different in comparison to other AFe compounds. MnFe-based PBAs are of particular interest due to their temperature- and photo-induced phase transitions; for example, Rb[MnFe(CN)$_6$] undergoes a structural transition from cubic $\mathrm{F4\overline{3}m}$ to tetragonal $\mathrm{I\overline{4}m2}$, driven by charge transfer from Mn$^{2+}$ to Fe$^{3+}$ ions, which induces concomitant structural and magnetic changes \cite{ohkoshi2005temperature}.

\section{Experimental methods}
\paragraph{Synthesis.}
The PBAs were synthesized by a co-precipitation route from aqueous solutions of the constituent metal salts and hexacyanometallate precursors, as detailed in Ref.~\cite{n2022toward}. When the synthesis is realized in presence of alkali cations (Rb), the alkali cation may be trapped in the interstitial sites of the structure. All syntheses were performed by the ICMMO synthesis platform; the full synthesis protocol (precursor concentrations, volumes, washing and drying procedure), together with the EDS/TGA compositional analysis, is provided in the Supplementary Material \cite{Supplement}.

\paragraph{Sample characterization and magnetization.}
Magnetization of synthesized powders was measured with a Quantum Design MPMS3 (LPS, Orsay) using standard zero-field-cooled (ZFC) and field-cooled (FC) protocols. In the paramagnetic regime, the magnetic susceptibility of all compounds was analyzed using the Curie--Weiss law to extract the Weiss temperature $\Theta$ and the effective moment $\mu_{\rm eff}$; the corresponding formulas, individual fits, and full ZFC/FC data sets are provided in the Supplementary Material \cite{Supplement}. For each compound, the ordering temperature $T_C$ derived from neutron diffraction (Sec.~\ref{sec:magref}) closely matches that obtained from bulk magnetization, confirming the robustness of the magnetic phase transitions reported in this work.

\textbf{Neutron diffraction}
Neutron diffraction was performed at D1B at ILL with a wavelength of 2.52 \AA., with orange cryostat or cryofurnace allowing to study the temperature range from 2K to 450K.  Neutron diffraction as function of magnetic field was performed using a 5 T magnetic field. The powder sample was loaded in the sample holder and
tamped down with a metallic rod to obtain a homogeneous packing within the sample holder; no controlled or quantified pressure was applied. A cadmium sheet was rolled and shaped into a cylindrical form and placed on top of the sample to avoid the displacement of the crystallites within the sample holder due to high magnetic field. The neutron diffraction was performed at 0 and 90 degrees, to check the possible grain orientation due to magnetic field.

\section{Structural role of vacancies and water}
\label{sec:structure}
\subsection{Vacancy-related diffuse scattering}

To establish the crystallographic framework common to the whole series before turning to its magnetic properties, we first examine the room-temperature neutron diffraction patterns of the bimetallic PBAs. In the neutron diffraction patterns of all investigated PBA compounds, a broad diffuse feature is systematically observed at low $2\theta$ angles. This feature has been attributed either to correlated structural disorder, such as vacancy–water site correlations \cite{franz2004crystalline}, or to short-range non-magnetic correlations. This unindexed reflection has been associated with partially correlated oxygen atoms; owing to the small number of oxygen atoms contributing to this feature, it is not visible in corresponding X-ray diffraction patterns \cite{kumar2007variation}.

\begin{figure}[!htb]
    \includegraphics[width=0.45\textwidth]{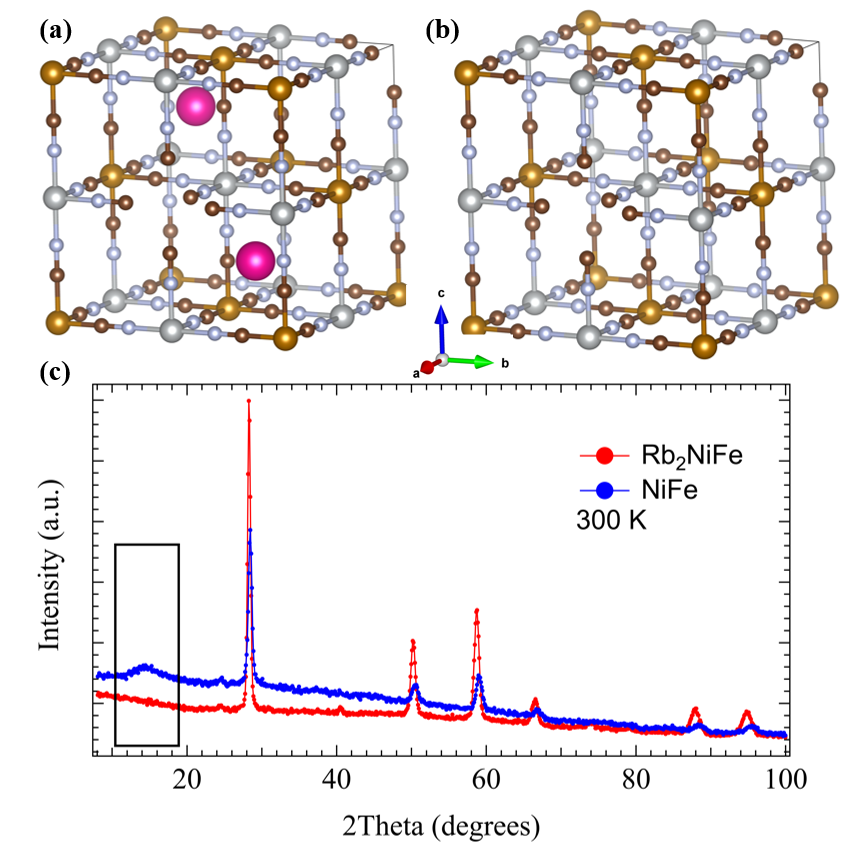}
    \caption{Crystal structure layouts of the bimetallic Prussian blue analogues: (a) the cation-stabilized Rb$_2$NiFe framework in the $\mathrm{F\overline{4}3m}$ space group and (b) NiFe crystallizing in the $\mathrm{Fm\overline{3}m}$ space group. For clarity, the H$_2$O molecules are omitted from both structural representations. (c) Room-temperature (300~K) neutron diffractograms of NiFe and Rb$_2$NiFe. The broad diffuse scattering feature (bump) at low $2\theta$ angles, associated with structural vacancies and coordinated water, is visibly absent in the Rb$_2$NiFe compound.}
 \label{fig1}
\end{figure}

To clarify the origin of this anomaly, we performed comparative neutron diffraction measurements on NiFe and Rb$_2$NiFe (Fig.~\ref{fig1}). In Rb$_2$NiFe, the Ni-NC-Fe framework is stabilized by interstitial Rb$^{+}$ cations, which reduce the available space for zeolitic water molecules and for vacancies. The absence of the broad low-$2\theta$ feature in Rb$_2$NiFe (Fig.~\ref{fig1}(c)) demonstrates that it predominantly originates from vacancy-related correlations. Moreover, the reduced hydrogen-like incoherent background in Rb$_2$NiFe indicates a lower interstitial water content. We can therefore conclude that this feature, common to all bimetallic PBAs of the Fm$\overline{3}$m family, is an intrinsic signature of the disordered vacancies and their associated water molecules, rather than of any structural or magnetic long-range order.

\subsection{Structural response to dehydration}
\label{sec:dehydration_structure}
\begin{figure}[!htb]
    \includegraphics[width=0.45\textwidth]{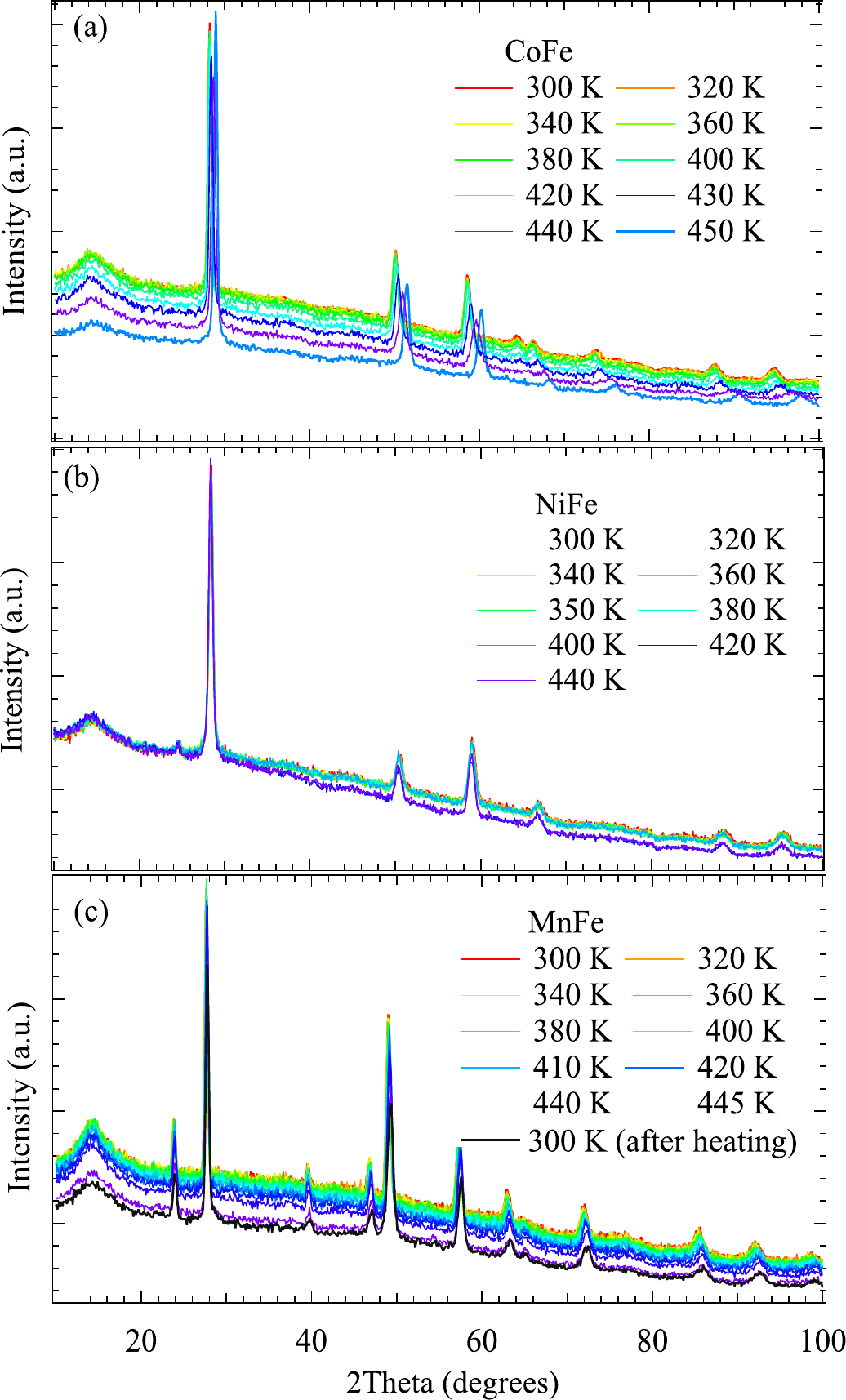}
    \caption{High-temperature neutron diffraction patterns for the AFe-based PBA compounds: (a) CoFe, (b) NiFe, and (c) MnFe. The data highlight the structural evolution and the systematic reduction of the hydrogen-induced incoherent scattering background upon heating/dehydration.}
    \label{figDehydration}
\end{figure}

Having established the structural signature of vacancies and water at 300~K, we now turn to the behavior of these compounds upon dehydration, both at elevated temperature (2--450~K) and after \textit{ex-situ} dehydration at low temperature. We performed temperature-dependent neutron diffraction measurements on AFe-based PBAs containing different 3$d$ transition-metal ions at the A site (Fig.~\ref{figDehydration}).

\begin{figure}[!htb]
    \includegraphics[width=0.45\textwidth]{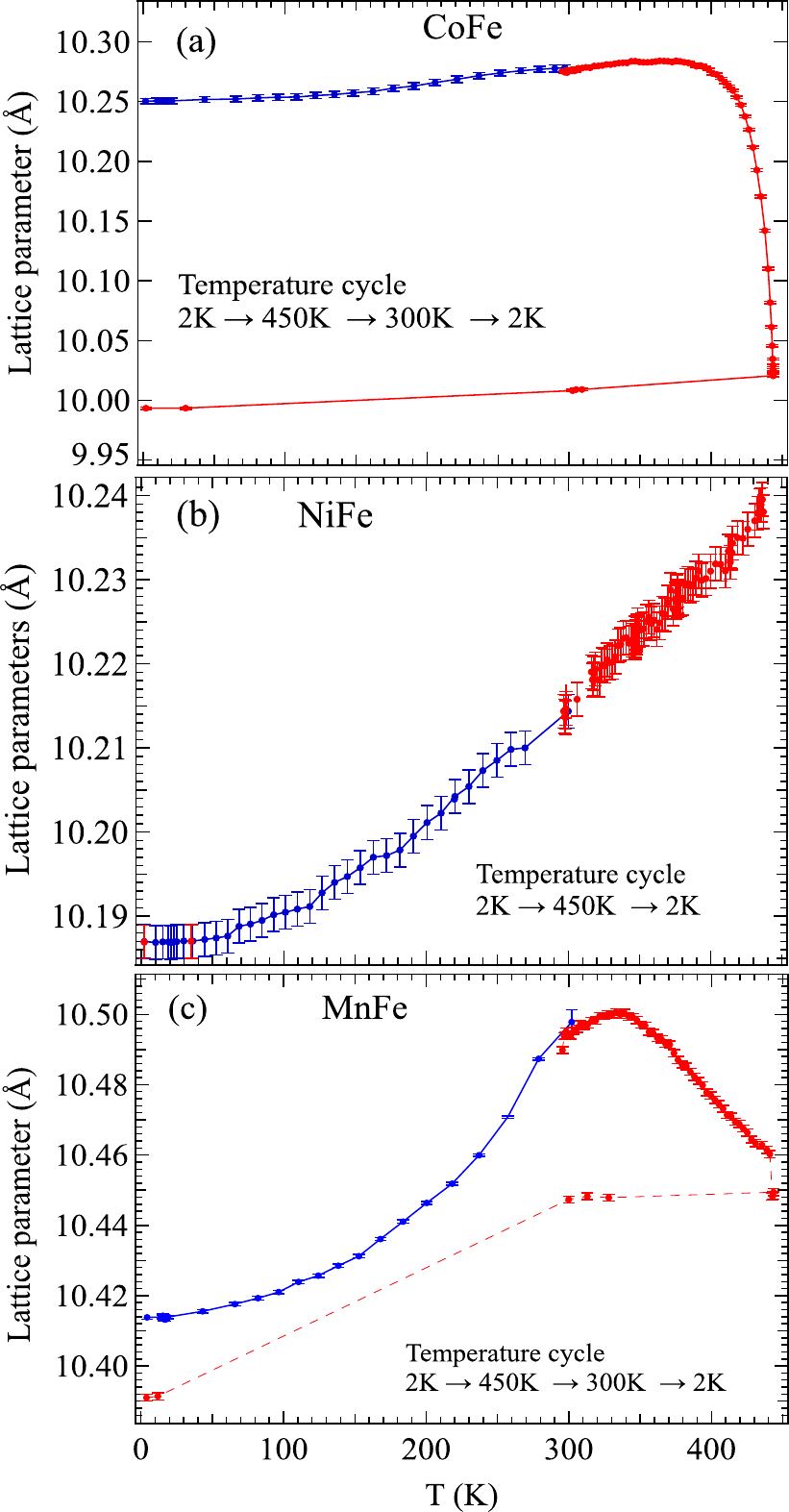}
    \caption{Temperature dependence of the lattice parameters for MnFe, CoFe, and NiFe. A continuous tuning from positive to negative thermal expansion coefficients is achieved upon heating for both MnFe and CoFe. Upon cooling back to 300~K after the full 2--450~K thermal cycle (i.e., in the dehydrated state), CoFe exhibits a structural contraction of 2.5\,\% relative to its initial hydrated state, while MnFe shows a minor reduction of 0.4\,\%. In contrast, NiFe displays a monotonic lattice expansion over the entire range.}
    \label{figLat_fT}
\end{figure}

In cyanide-based frameworks, the coefficient of thermal expansion (CTE) is strongly influenced by guest content and can span positive, negative, or near-zero values depending on pore occupancy. At elevated temperatures, the loss of interstitial and coordinated water generally destabilizes the framework. Among the investigated compounds, NiFe remains structurally stable throughout the measured temperature range, likely owing to its relatively small lattice parameter, which enhances confinement of water molecules. In contrast, CoFe and MnFe display a continuous crossover from positive to negative thermal expansion (NTE) upon heating (Fig.~\ref{figLat_fT}), consistent with a similar continuous evolution of the CTE previously reported for CoFe PBAs upon progressive CO$_2$ adsorption \cite{auckett2018continuous}. The interpretation of the high-temperature behavior of CoFe is nonetheless complicated by the possible occurrence of a charge-transfer process (Co$^{2+}$/Fe$^{3+}$ $\leftrightarrow$ Co$^{3+}$/Fe$^{2+}$) in addition to water loss; disentangling these two contributions will require dedicated experiments.

Engineering zero thermal expansion (ZTE) in PBAs has been widely pursued through controlled incorporation of guest ions and H$_2$O molecules into interstitial sites. In this context, Rb$_2$NiFe exhibits an approximately fourfold reduction in CTE compared to NiFe, attributed to interstitial Rb$^{+}$ ions acting as lattice anchors that stiffen the framework -- in contrast to reports where framework cation substitution (e.g., Ga$^{3+}$, which is part of the coordination polymer itself) instead enhances PTE by suppressing the NTE-driving transverse modes \cite{gao2018tunable}. The temperature dependence of the lattice parameters for MnCr and Rb$_2$NiFe, together with an extended literature comparison of the extracted CTE values, is provided in the Supplemental Material \cite{Supplement}.

The linear thermal expansion coefficient is defined as
\[\alpha = \frac{1}{a}\frac{da}{dT} = \frac{d\ln a}{dT},\]
and the extracted values in the respective temperature regimes are summarized in Table~\ref{table3}, consistent with previous reports \cite{adak2011thermal}. Across the A-site series from Mn to Ni, the magnitude of the positive CTE decreases with increasing atomic number, indicating a correlation between thermal expansion behavior and cation size, as also observed in M[Pt(CN)$_6$] frameworks \cite{chapman2006compositional}. The physical origin of NTE in PBAs is well established: thermally activated transverse vibrations of the bridging cyanide ligands draw adjacent metal centers closer together, as demonstrated for Zn[Pt(CN)$_6$] \cite{goodwin2005guest}, owing to the inherent flexibility of the --M--C$\equiv$N--M-- linkages (M = metal).

\begin{table*}[htb!]
\centering
\begin{tabular}{c c c c c c c}
 \hline
Compound & a (4 K) Å  & a (300 K) Å  & a (450 K) Å  & $\alpha$ (4-300 K)  & $\alpha$ (300-450 K) & Ref\\
\hline
MnFe &  10.42(1)  & 10.49(1) & 10.46(1) & 29$\times$ 10$^{-6}$ K$^{-1}$ & - 20$\times$ 10$^{-6}$ K$^{-1}$ & \cite{adak2011thermal}\\
CoFe & 10.27(1)  & 10.30(1) &  10.03(1) & 8.5$\times$ 10$^{-6}$ K$^{-1}$  & - 1.5$\times$ 10$^{-4}$ K$^{-1}$  & \cite{adak2011thermal}\\
NiFe & 10.187(4)  & 10.22(1) & 10.24(1)  & 11$\times$ 10$^{-6}$ K$^{-1}$ & 15$\times$ 10$^{-6}$ K$^{-1}$  & \cite{adak2011thermal}\\
MnCr & 10.65(1) & 10.73(1)* & -& 12$\times$ 10$^{-6}$ K$^{-1}$ &N/A & \cite{adak2024thermal}\\
Rb$_2$NiFe & 10.27(1)  & 10.28(1) & - & 3.1$\times$ 10$^{-6}$ K$^{-1}$ & N/A & \cite{matsuda2009universal} \\
\hline
\end{tabular}
\caption{Lattice parameters at 4~K, 300~K (except for MnCr at 225~K, marked by *), and 450~K, alongside the linear thermal expansion coefficients ($\alpha$) extracted in the low- and high-temperature regimes. All compounds exhibit positive thermal expansion (PTE) at low temperatures (4--300~K), whereas specific compounds undergo a crossover to negative thermal expansion (NTE) in the high-temperature region (300--450~K). The pronounced NTE observed in CoFe correlates with the reduction of its magnetic ordering temperature.}
\label{table3}
\end{table*}

\begin{figure}[!htb]
    \centering
    \includegraphics[width=0.5\textwidth]{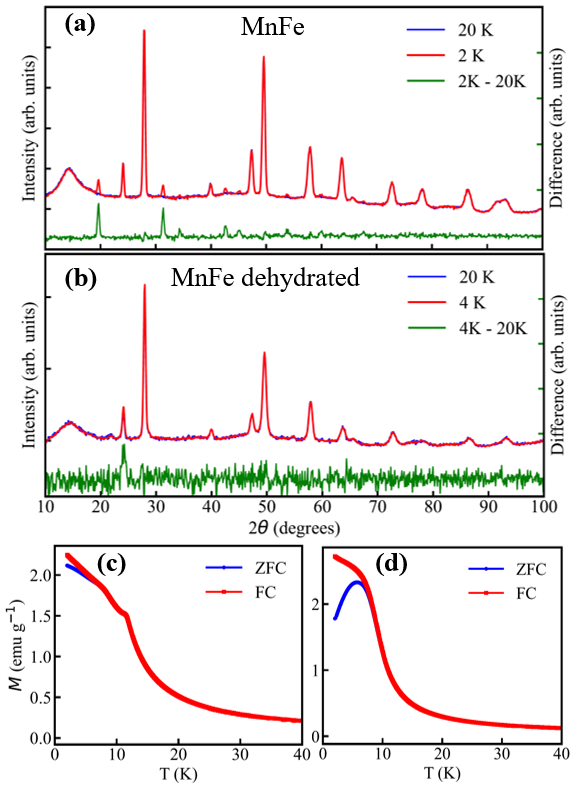}
    \caption{(a) Neutron diffraction patterns at 20~K and 4~K, together with their difference, for non-dehydrated MnFe; (b) corresponding data for the dehydrated sample. The difference pattern of the non-dehydrated sample reveals distinct magnetic Bragg peaks, whereas in the dehydrated sample the magnetic contribution coincides with the nuclear peaks, indicating a modified magnetic structure. The corresponding magnetization curves under ZFC and FC protocol at 500 Oe for the non-dehydrated and dehydrated samples are shown in (c) and (d), respectively.}
 \label{figMnFe}
\end{figure}

\begin{figure}[!htb]
    \centering
    \includegraphics[width=0.5\textwidth]{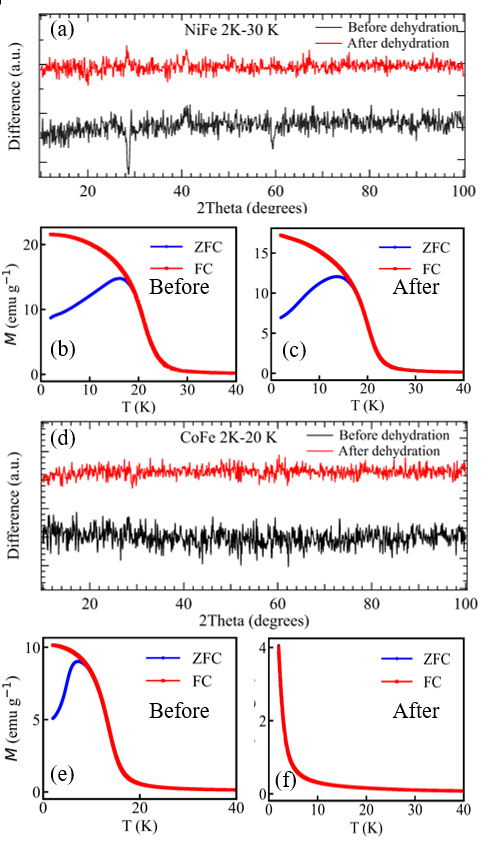}
    \caption{Effect of dehydration: (a) The difference of neutron diffraction data of NiFe at 2 K and 30 K before and after dehydration. (b,c) Magnetization under ZFC and FC protocol at 500 Oe of NiFe before and after dehydration. (d) CoFe: Difference of neutron diffractograms at 2 K and 20 K before and after dehydration. (e, f) ZFC and FC Magnetization at 500 Oe before and after dehydration, respectively.}
    \label{figCoFe_NiFe}
\end{figure}

Complementary to the high-temperature study, we also compared the low-temperature (2--4~K) neutron diffraction patterns of MnFe, CoFe, and NiFe before and after a full dehydration cycle (i.e., after being heated to 450~K and cooled back down). Substantial modifications in the diffraction patterns are observed for CoFe and MnFe, whereas NiFe shows only minor changes, primarily reflected in the incoherent background (Fig.~\ref{figMnFe} and Fig.~\ref{figCoFe_NiFe}), directly corroborating the crucial role of interstitial water in the emergence of the low-angle diffuse scattering feature discussed above. The magnetic consequences of this structural dehydration are discussed for MnFe in Sec.~\ref{sec:mnfedehydration}, and for CoFe and NiFe in the Supplementary Material \cite{Supplement}.

These results demonstrate that the presence of H$_2$O plays a crucial role in governing both the structural and magnetic properties of these compounds. As dehydration drives a reduction of the lattice parameter, the 3d ions are brought into closer proximity, thereby enhancing the magnetic exchange interactions and stabilizing a distinct magnetic ground state, consistent with the exchange interaction pathways in PBAs via the cyanide bridges: once the structural framework is compromised, the magnetic order is correspondingly affected \cite{zhang2022lithiated, lu2006tuning}. The microscopic connection between lattice contraction and the response of the CN bonds to dehydration remains to be explored further.

\subsection{Structural classification of the compounds}
All bimetallic systems MnFe, CoFe, NiFe, and MnCr crystallize in the cubic Fm$\overline{3}$m structure (space group No.~225), whereas Rb$_2$NiFe (also considered bimetallic as the Rb$^+$ ion occupies interstitial sites and is not a framework transition-metal ion) adopts the non-centrosymmetric cubic F$\overline{4}$3m structure (space group No.~216), consistent with earlier reports \cite{tokoro2011novel}. The lattice parameters at 300~K and space groups of all compounds are summarized in Table~\ref{table1}, together with the shorthand notation (MnFe, CoFe, NiFe, MnCr, Rb$_2$NiFe) used throughout this article to refer to each compound. Notably, all neutron diffraction patterns exhibit a pronounced incoherent background originating from hydrogen, in line with previous observations \cite{temleitner2015neutron}. The corresponding magnetic ground states and ordering temperatures, in excellent agreement with the values independently derived from bulk magnetization, are discussed in Sec.~\ref{sec:magref} below.

\begin{table*}[htb!]
\centering
\begin{tabular}{c c c c}
 \hline
Composition & abbreviation & Symmetry & a (\AA) \\

\hline
Co$_4$[Fe(CN)$_6$]$_{2.7}$ & CoFe & $\mathrm{Fm\overline{3}m}$& 10.30(1)  \\
Mn$_4$[Fe(CN)$_6$]$_{2.7}$ & MnFe & $\mathrm{Fm\overline{3}m}$&  10.49(1) \\
Ni$_4$[Fe(CN)$_6$]$_{2.7}$ & NiFe & $\mathrm{Fm\overline{3}m}$&  10.22(1) \\
Rb$_2$Ni$_4$[Fe(CN)$_6$]$_{3.3}$ & Rb$_2$NiFe & $\mathrm{F4\overline{3}m}$ &10.28(1)  \\
Mn$_4$[Cr(CN)$_6$]$_{2.7}$ & MnCr & $\mathrm{Fm\overline{3}m}$& 10.73(1)*  \\
\hline
\end{tabular}
\caption{Composition, shorthand notation, space group, and lattice parameter determined by Rietveld refinement of neutron diffraction data at 300~K (except for MnCr, measured at 225~K and indicated by *).}
\label{table1}
\end{table*}

\section{Magnetic reference states across the PBA series}
\label{sec:magref}
\subsection{Ordering temperatures and bulk response}
Low-temperature neutron diffraction measurements were carried out to determine the magnetic ordering temperatures and ground-state magnetic structures of the investigated PBA compounds; no temperature-induced structural phase transition was observed between 300~K and 2~K. The magnetic ordering temperatures were extracted from the temperature dependence of the integrated intensity of the magnetic Bragg peaks. For every compound, the resulting neutron order parameter is in excellent agreement with the bulk magnetization data, with both probes showing a clear onset at the same temperature.

Most compounds of the PBA family investigated here exhibit a ferrimagnetic (uncompensated antiferromagnetic) ground state, with the notable exception of the isostructural ferromagnetic NiFe and Rb$_2$NiFe. Since ferri- and ferromagnetic materials show an increase in magnetization with increasing applied magnetic field, a direct comparison among the different compounds of this family requires measurements at a fixed field; the magnetization values measured at an applied field of 500~Oe, together with the Curie--Weiss parameters and effective moments, are summarized in Table~\ref{table2}.

For the bimetallic compounds, the paramagnetic susceptibility is, to a good approximation, the sum of the independent Mn (or Co, Ni) and Fe (or Cr) sublattice contributions, so that only the Curie constants -- not the individual effective moments -- are additive: $C_{\rm tot} = n_A C_A + n_B C_B$, and therefore
\[
\mu_{\mathrm{eff,tot}} = \sqrt{n_A\,\mu_{\mathrm{eff},A}^2 + n_B\,\mu_{\mathrm{eff},B}^2}\\
\]
\[
\mu_{{\rm eff},i} = g\sqrt{S_i(S_i+1)},
\]
with $n_A$, $n_B$ the number of A- and B-site ions per formula unit. The spin-only estimates quoted below use this combination rule, together with the electronic states established from the neutron and Curie--Weiss analysis: Mn$^{2+}$ ($3d^5$, high-spin, $S=5/2$), Fe$^{3+}$ ($3d^5$, low-spin, $S=1/2$), and Cr$^{3+}$ ($3d^3$, $S=3/2$, the only accessible spin state for this configuration in an octahedral field), all with $g=2$.

\begin{figure}[!htb]
    \centering
    \includegraphics[width=0.45\textwidth]{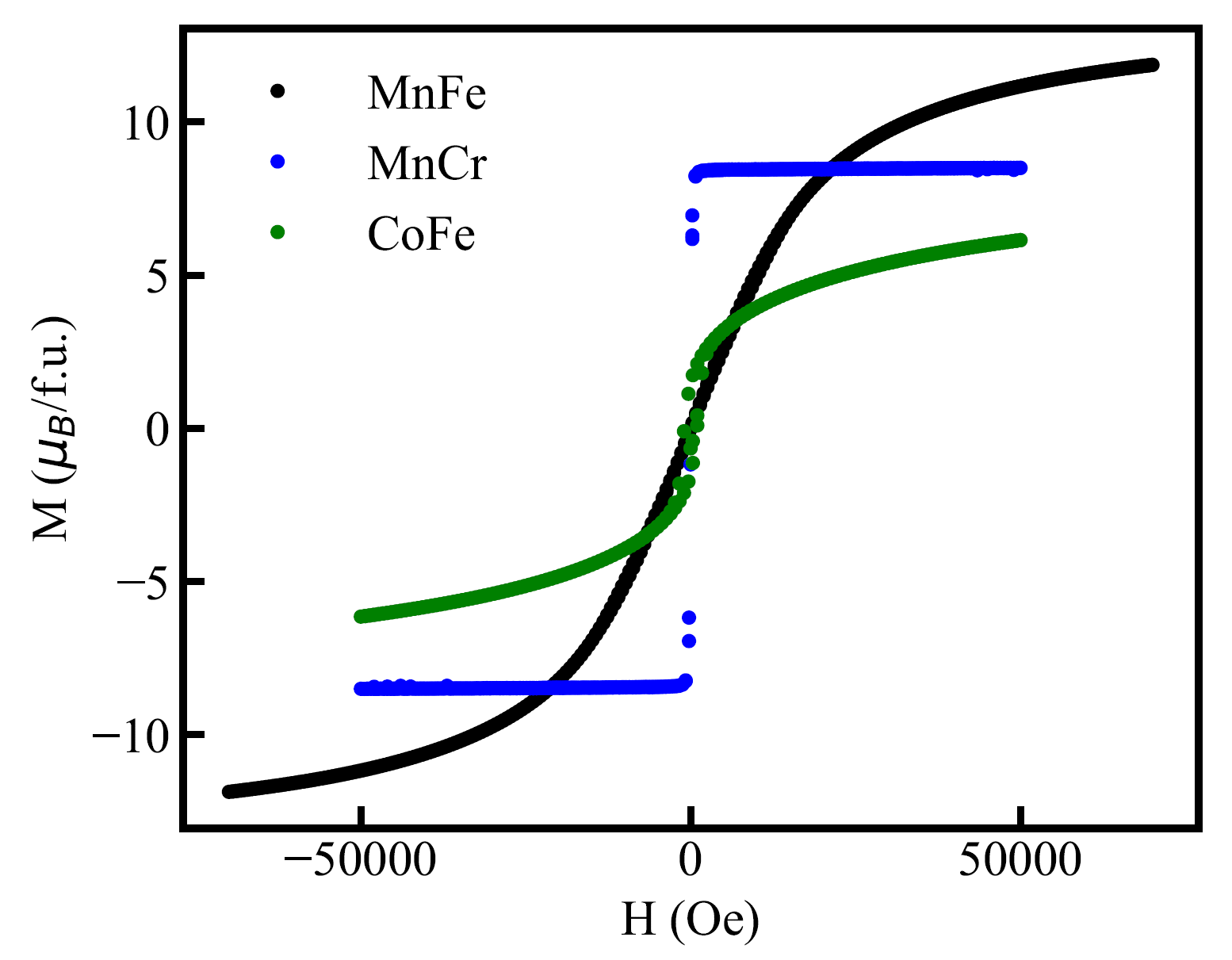}
    \caption{Field-dependent magnetization ($M$--$H$) curves measured at 4~K for the studied PBA series. MnFe displays a strong magnetic anisotropy preventing saturation up to 7~T, whereas MnCr exhibits a soft ferromagnetic-like response with rapid saturation starting at approximately 0.1~T.}
 \label{figMsPBA}
\end{figure}

Magnetization as a function of the applied magnetic field was measured at low temperature ($\sim 4$~K), well below the magnetic ordering temperatures (Fig.~\ref{figMsPBA}). For most of the investigated PBAs, such as MnFe, the magnetization does not reach complete saturation even at fields up to 7~T, indicating the presence of strong magnetic anisotropy or complex exchange interactions. In contrast, the MnCr compound exhibits a clear saturation below $\sim 0.1$~T. Temperature and field-dependent magnetization measurements show that the $\mu_{\rm eff}$ values in the paramagnetic regime are larger for MnFe and MnCr compared to the other compounds in the family, consistent with their strong magnetic sublattice moments; all compounds exhibit typical ferri- or ferromagnetic behavior. Additionally, for MnFe, an irreversibility temperature $T_{\rm irr} \sim 20$~K is observed, below which the ZFC and FC curves start to bifurcate.

\begin{table*}[htb!]
\centering
\footnotesize
\begin{tabular}{c c c c c c c c}
 \hline
Compound & Magnetic state & $T_C$ (K) & $\Theta$ (K)  &$\mu_{\rm eff}$ (CW) & $\mu_{\rm eff}$ (spin-only) & $M_{s}$ ($\mu_B$/f.u.) & $M_{net}$ (neutron, 0 T) \\ \hline
CoFe & FiM & 14 & -20 & 9.04 $\mu_B$/f.u. & -- & 4.62 (5 T) & -- \\
MnFe & FiM & 12 & -10 & 10.41 $\mu_B$/f.u  & 12.17 $\mu_B$/f.u.  & 9.91 (5 T) & 12.48 $\mu_B$/f.u.\\
NiFe & FM & 21 & 26 & 5.90 $\mu_B$/f.u. & -- & 5.37 (5 T) & -- \\
Rb$_2$NiFe & FM & 26 & 29 & 7.46 $\mu_B$/f.u. & -- & -- & -- \\
MnCr & FiM & 64 & -34  &10.15 $\mu_B$/f.u. & 13.44  $\mu_B$/f.u. & 9.1 (5 T) & 10.86  $\mu_B$/f.u.\\
\hline
\end{tabular}
\caption{Magnetic ground state (FM: ferromagnetic, FiM: ferrimagnetic), ordering temperature ($T_C$), Curie--Weiss parameters, and moments for all investigated PBAs. Curie--Weiss fitting was performed at 500~Oe (150--300~K for MnFe, 200--300~K for MnCr; see Supplementary Material for the other compounds). $M_{net}$ (neutron, 0~T) is only reported for MnFe and MnCr, the only two compounds for which the magnetic contribution was strong enough for a full magnetic structure refinement (Sec.~\ref{sec:magstruct0T}); entries left blank (--) were not separately tabulated in this work (see Supplementary Material for the raw Rb$_2$NiFe Curie--Weiss data).}
\label{table2}
\end{table*}

\subsection{Magnetic structures at zero field}
\label{sec:magstruct0T}
Below the magnetic ordering temperature, an additional magnetic contribution appears superimposed on the nuclear Bragg peaks for all compounds except MnFe. In MnFe, the emergence of new magnetic reflections with propagation vector $\mathbf{k} = (1,0,0)$ is observed, immediately distinguishing it from the rest of the family and singling it out for a dedicated discussion in Sec.~\ref{sec:mnfe}. The magnetic contributions in CoFe, NiFe, and Rb$_2$NiFe are comparatively weaker than in MnFe and MnCr, which precludes reliable magnetic structure determination for these compounds. Consequently, only MnCr and MnFe exhibit well-resolved magnetic diffraction patterns suitable for full magnetic structure refinement.

The magnetic structures were determined by Rietveld refinement using the FullProf suite~\cite{Rodriguez1993} on difference data between temperatures well below and well above $T_C$. For both propagation vectors observed in this study, the compatible magnetic space groups were enumerated from the parent structure Fm$\overline{3}$m (\#225) using MAXMAGN~\cite{perez2015symmetry, gallego2012magnetic}; the full symmetry analysis is provided in the Supplementary Material \cite{Supplement}. In brief, for $\mathbf{k}=(0,0,0)$, of the twelve magnetic space groups compatible with the parent structure, only $I4/mm'm'$ (\#139.537, moment along a cubic axis) and $Im'm'm$ (\#71.536, moment along a face diagonal) allow a non-zero ordered moment.

\textbf{MnCr:} For the MnCr PBA compound, with propagation vector $\mathbf{k} = 0$, the $I4/mm'm'$ solution provides the better agreement with the data among the two candidate magnetic space groups, showing that the ordered moments are constrained along a cubic crystallographic axis. The Rietveld refinement yields ordered magnetic moments of Mn = $+4.90\pm 0.09\mu_B$ and Cr = $-2.63\pm 0.14\mu_B$, consistent with the high-spin states of Mn$^{2+}$ ($S=5/2$) and Cr$^{3+}$ ($S=3/2$). The neutron diffraction refinement yields a ferrimagnetic net moment of $M_{\rm net} \approx 12.50\,\mu_B$ per formula unit, and magnetization at 5~T gives $M_{\rm s} \approx 9.1\,\mu_B$/f.u., about 73\% of the neutron-derived value. The magnetic structure of MnCr is presented in the supplementary material \cite{Supplement}.

The systematic discrepancy between the magnetic moment derived from bulk magnetic susceptibility and that obtained from neutron diffraction ($\sim$75\%) originates from the fact that neutron diffraction determines the ordered moment directly on the magnetic site, since the nuclear diffraction pattern provides an internal normalization that fixes the scale factor. In contrast, the moment derived from bulk susceptibility depends on normalization by the sample mass, which is less accurate as it depends not only on the compound stoichiometry but, more critically, on the water content — a factor from which the neutron-derived moment is free. The "magnetic mass" is therefore overestimated due to this water content, which would then typically account for about 25\% of the total sample mass, explaining the difference between the two determinations. This is directly confirmed by thermogravimetric analysis (TGA): the interstitial and coordination water content determined by TGA amounts to $\sim$16.5~H$_2$O per formula unit, i.e.\ $\sim$25--27\% of the total sample mass depending on the compound, in close quantitative agreement with the value inferred above (see Supplementary Material \cite{Supplement} for the EDS/TGA-derived stoichiometries and water contents).

\subsection{Field response of the reference compounds}
\label{sec:fieldref}
\begin{figure}[!htb]
    \includegraphics[width=0.45\textwidth]{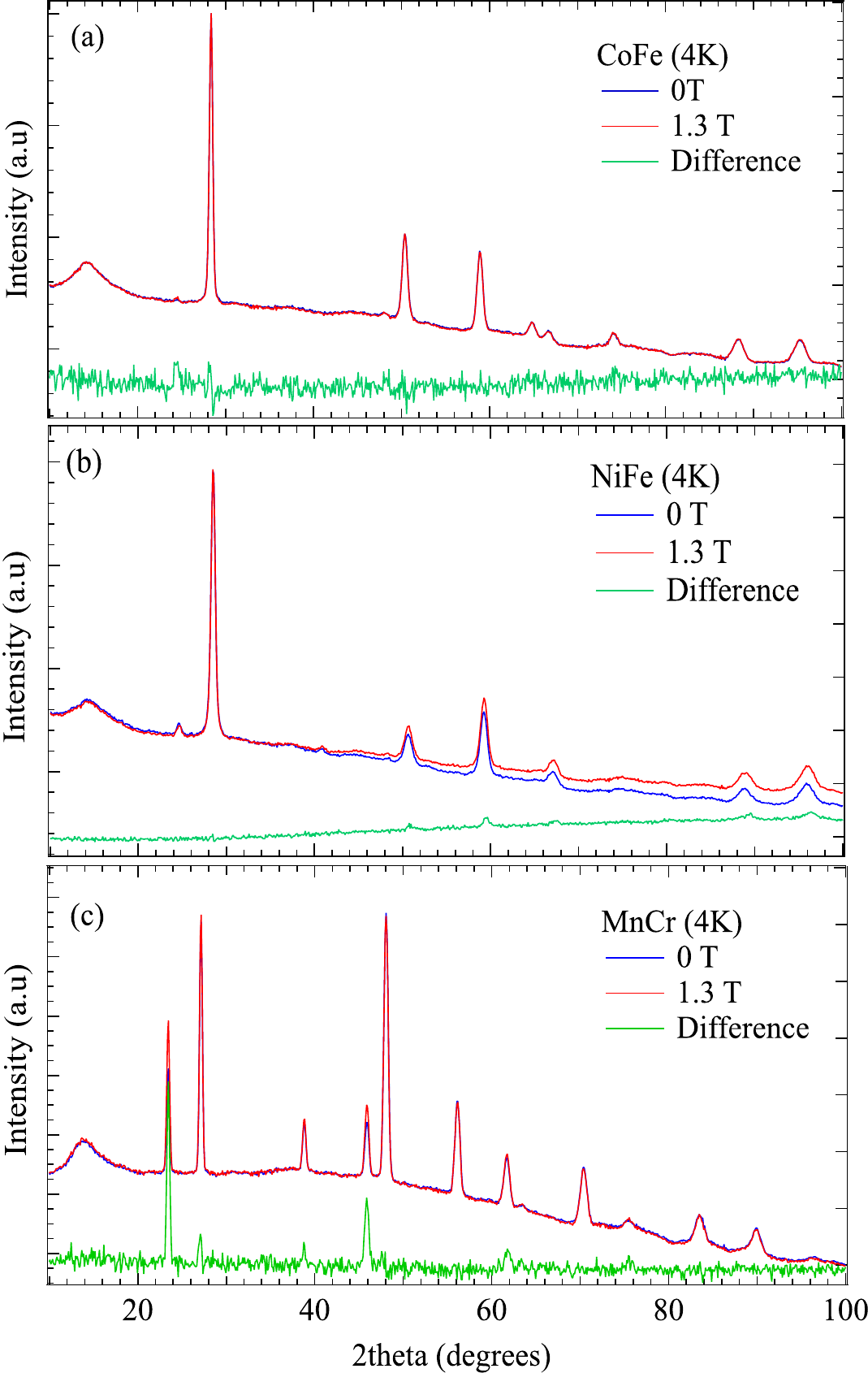}
    \caption{The Neutron diffraction under magnetic field at 4 K, for (a) CoFe, (b) NiFe and (c) MnCr PBA compounds.}
    \label{figS12}
\end{figure}

We investigated the effect of an external magnetic field on the magnetic ground states of all studied PBA compounds by performing powder neutron diffraction measurements as a function of applied field within the magnetically ordered regime. No field-induced structural transitions were observed in any compound, as evidenced by the absence of modifications to the nuclear Bragg peaks. With increasing field, only a slight enhancement of the magnetic scattering intensity was observed for NiFe and CoFe (Fig.~\ref{figS12}), consistent with a progressive alignment of magnetic domains within an already established $\mathbf{k} = (0,0,0)$ ferrimagnetic structure. MnCr exhibits a more pronounced field-induced increase in magnetic intensity, also detailed in the Supplementary Material \cite{Supplement}. These behaviors of CoFe, NiFe and MnCr, essentially insensitive to the applied field, are in stark contrast to the unique response of MnFe described in the following section.

\section{MnFe: a switchable magnetic ground state}
\label{sec:mnfe}
\subsection{Zero-field frustrated structure}
Among the three magnetic space groups compatible with $\mathbf{k} = (1,0,0)$ and a non-zero total magnetic moment (enumerated via MAXMAGN in the Supplementary Material \cite{Supplement}), only $P_I4/mnc$ (\#128.410, moment along a cubic axis) provides a satisfactory fit to the Rietveld refinement of the MnFe zero-field difference data, demonstrating that the ordered Mn and Fe moments are constrained along a cubic crystallographic axis rather than along a face diagonal. The resulting magnetic structure is shown in Fig.~\ref{figS1}(c); the Rietveld refinement curves and fitting parameters are presented in the Supporting Material \cite{Supplement}.

The refined magnetic moments are Mn = $+3.52 \pm 0.05 \mu_B$, Fe = $-0.59\pm 0.07\mu_B$, confirming the high-spin (HS) state of Mn$^{2+}$ ($t_{2g}^3 e_g^2$) and low-spin (LS) state of Fe$^{3+}$ ($t_{2g}^5 e_g^0$). The net magnetic moment per formula unit, $M_{\rm net}$, in the ferrimagnetic state is estimated as:
\[ M_{\rm net} = \left| n_{\rm Mn} \, \mu_{\rm Mn}^{\rm ord} - n_{\rm Fe} \, \mu_{\rm Fe}^{\rm ord} \right| =  12.48\,\mu_B \]
The saturation magnetization, $M_{\rm s} = 9.91\,\mu_B$ (Fig.~\ref{figMsPBA}), is approximately 79\% of the net moment extracted from neutron diffraction, reflecting the presence of uncompensated local moments per formula unit in the ferrimagnetic ground state. Using the spin-only model for Mn$^{2+}$ ($3d^5$, high-spin, $S = 5/2$) and Fe$^{3+}$ ($3d^5$, low-spin, $S = 1/2$), and combining the two sublattices through their Curie constants ($n_{\rm Mn}=4$, $n_{\rm Fe}=2.7$ per formula unit; see Sec.~\ref{sec:magref}), one finds $\mu_{\rm eff} = \sqrt{4\times(2\sqrt{35/4})^2 + 2.7\times(2\sqrt{3/4})^2} \approx 12.17\,\mu_B$/f.u., in good agreement with the neutron result. Unlike CoFe, NiFe, and MnCr (Sec.~\ref{sec:fieldref}), the ordered Mn moment in MnFe is thus comparatively far from its expected saturated value at zero field, as further explored below.

\subsection{Field-induced transition}

\begin{figure*}[!htb]
    \centering
    \includegraphics[width=0.95\textwidth]{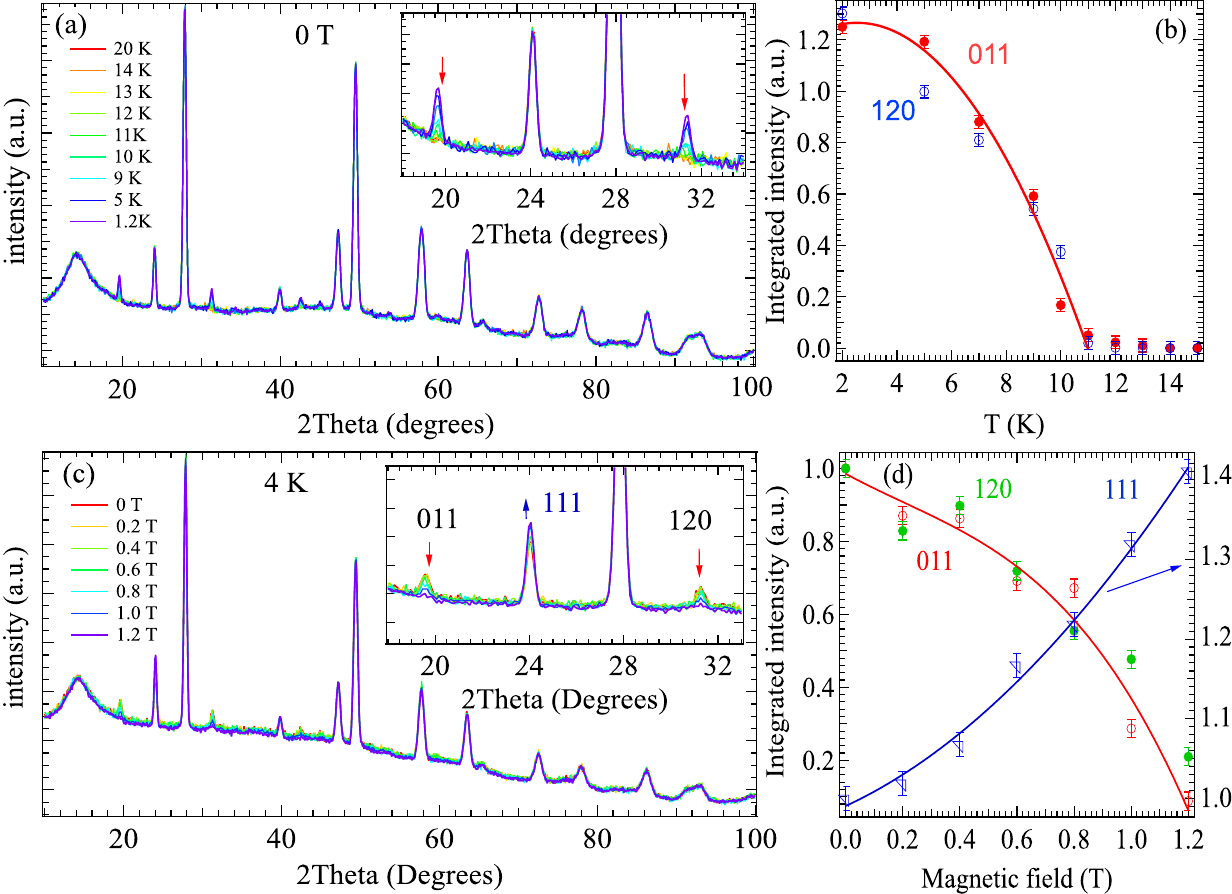}
    \caption{(a) Temperature-dependent neutron diffraction profiles of MnFe collected at 0~T. (b) Integrated intensity of the magnetic Bragg reflections as a function of temperature at zero applied field, outlining the magnetic order parameter. (c) Magnetic-field dependence of the neutron diffractograms recorded at 4~K. (d) Field-driven intensity evolution of the (011) and (120) magnetic reflections (left axis), characteristic of the low-field $\mathbf{k} = (1,0,0)$ phase, and of the (111) reflection (right axis), corresponding to the field-induced $\mathbf{k} = (0,0,0)$ collinear ferrimagnetic state. Solid lines are guides to the eye.}
    \label{figS10}
\end{figure*}

MnFe displays a markedly distinct field-dependent behavior, which is particularly noteworthy given its magnetic ground state characterized by a propagation vector $\mathbf{k} = (1,0,0)$, in contrast to the $\mathbf{k} = (0,0,0)$ order found in all other compounds. As shown in Fig.~\ref{figS10}(c,d), the intensity of the magnetic Bragg peaks associated with $\mathbf{k} = (1,0,0)$ decreases continuously with increasing field and vanishes completely above $B_c = 1.3$~T. Simultaneously, an increase in intensity at the nuclear Bragg peak positions demonstrates a transfer of magnetic spectral weight toward $\mathbf{k} = (0,0,0)$. Over this same field range, the purely nuclear reflections remain invariant in intensity, indicating the absence of significant field-induced grain movement or reorientation within the powder sample. Magnetization measurements as a function of applied field further confirm this singular behavior: while the applied field has almost no impact on CoFe and NiFe, MnFe exhibits a smearing of the magnetic ordering temperature with increasing field (Supplementary Material \cite{Supplement}).

At 5~T, the diffraction pattern of MnFe is fully consistent with the $I4/mm'm'$ magnetic space group identified above for $\mathbf{k} = 0$ structures --- the same solution found for MnCr at zero field. This confirms that, above $B_c$, the ordered Mn and Fe moments of MnFe are likewise constrained along a cubic crystallographic axis, and that the field-driven collinear ferrimagnetic state of MnFe is symmetry-equivalent to the zero-field ground state of the other $A$Fe and MnCr compounds of the family. Rietveld refinement of the field-dependent neutron diffraction data yields the evolution of the Mn and Fe ordered moments as a function of applied field.

\begin{figure}[!htb]
    \includegraphics[width=0.5\textwidth]{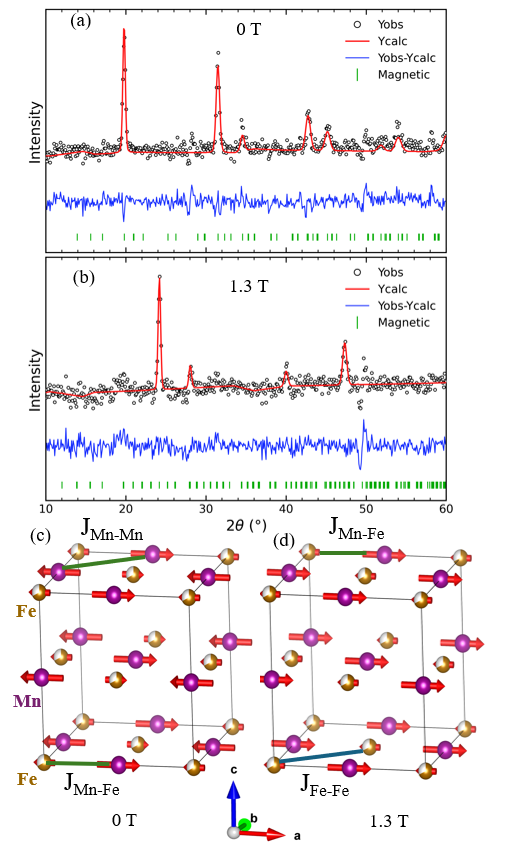}
    \caption{Magnetic field induced spin reorientation from the frustrated state ($\mathbf{k}=(1,0,0)$) to the collinear ferrimagnetic state ($\mathbf{k}=(0,0,0)$). The nearest-neighbor and second-neighbor couplings are represented in blue and green, respectively. (a) Rietveld refinement of the MnFe at 0 T and (b) 5 T.   (c) MnFe magnetic structure at 0~T. (d) MnFe magnetic structure at 5~T, showing the ferrimagnetic ground state. Above 1.3~T, all nearest-neighbor Mn--Fe pairs are antiferromagnetically coupled and all second-nearest-neighbor pairs are ferromagnetically coupled.}
    \label{figS1}
\end{figure}

The resulting magnetic structure at 5~T corresponds to a fully collinear ferrimagnetic state (Configuration~II, $\mathbf{k} = (0,0,0)$), in which all Mn moments are aligned along $+\hat{a}$ and all Fe moments along $-\hat{a}$ (Fig.~\ref{figS1}). This field-induced spin reorientation from the frustrated $\mathbf{k} = (1,0,0)$ state to the collinear ferrimagnetic $\mathbf{k} = (0,0,0)$ state appears to be an intrinsic property of MnFe-based PBA compounds, as it has been observed in both bimetallic and trimetallic members of the family. A field-induced antiferromagnetic-to-ferrimagnetic transition was similarly reported in Rb$_{0.19}$Ba$_{0.3}$Mn$_{1.1}$[Fe(CN)$_6$]$\cdot$0.48H$_2$O~\cite{yusuf2012magnetic}, and a field-induced spin reversal was likewise observed in Cu$_{0.73}$Mn$_{0.77}$[Fe(CN)$_6$]$\cdot z$H$_2$O around 20~kOe~\cite{lahiri2016understanding}.

We note that grain-to-grain variability in the response to the applied field may partly contribute to the apparent reduction of the ordered moments. While the field-induced change in propagation vector is unambiguously established, the associated magnetic space groups represent the most probable solutions compatible with our powder data, which single-crystal measurements would help confirm.

\subsection{Exchange model}
The field-driven switching between the two magnetic configurations of MnFe provides a direct experimental handle on the competing magnetic exchange interactions at play. To extract quantitative information on the exchange interactions, we consider a minimal spin Hamiltonian defined on the face-centered cubic (FCC) unit cell containing four Mn and four Fe atoms. For simplicity, the exchange bonds are enumerated on this idealized, fully-occupied FCC framework; the effect of the Fe vacancies (Fe occupancy $\approx$2.7/4, see Supplementary Material \cite{Supplement} for the EDS-derived stoichiometry) on the local exchange topology is neglected in this minimal model. The Hamiltonian includes nearest-neighbor Mn--Fe interactions and second-neighbor Mn--Mn and Fe--Fe interactions:
\begin{equation}
\begin{split}
\mathcal{H} & =
J_{\mathrm{Mn\text{-}Fe}} \sum_{\langle i,j \rangle}^{12} \mathbf{S}^{\mathrm{Mn}}_i \cdot \mathbf{S}^{\mathrm{Fe}}_j
+ J_{\mathrm{Mn\text{-}Mn}} \sum_{\langle i,j \rangle}^{6}  \mathbf{S}^{\mathrm{Mn}}_i \cdot \mathbf{S}^{\mathrm{Mn}}_j \\
& + J_{\mathrm{Fe\text{-}Fe}} \sum_{\langle i,j \rangle}^{6} \mathbf{S}^{\mathrm{Fe}}_i \cdot \mathbf{S}^{\mathrm{Fe}}_j
- \mathbf{B} \cdot \left( \sum_{i=1}^{4} \mathbf{S}^{\mathrm{Mn}}_i + \sum_{i=1}^{4} \mathbf{S}^{\mathrm{Fe}}_i \right),
\label{eq:hamiltonian}
\end{split}
\end{equation}
where the last term is the Zeeman coupling to the applied field. With this sign convention, a positive $J$ denotes an antiferromagnetic coupling and a negative $J$ a ferromagnetic one, consistently for every bond. Consistent with the observed magnetic ground state, we assume $J_{\mathrm{Mn\text{-}Fe}} > 0$ and $J_{\mathrm{Mn\text{-}Mn}},\, J_{\mathrm{Fe\text{-}Fe}} > 0$ (all three couplings antiferromagnetic in this convention). The observed partially frustrated $\mathbf{k}=(1,0,0)$ ground state then reflects the well-known geometric frustration of the antiferromagnetically coupled Mn face-centered-cubic sublattice, which cannot simultaneously satisfy all six second-neighbor Mn--Mn bonds; combined with the antiferromagnetic Mn--Fe coupling, this frustration stabilizes the observed canted structure over the fully collinear configuration.

We consider two competing collinear spin configurations. Configuration~I is the partially frustrated state with propagation vector $\mathbf{k} = (1,0,0)$, observed experimentally at zero field, in which three Mn moments point along $+\hat{a}$ and one along $-\hat{a}$, with Fe moments locally antiparallel to their Mn neighbors. Configuration~II is the fully collinear ferrimagnetic state with $\mathbf{k} = (0,0,0)$, in which all Mn moments are aligned along $+\hat{a}$ and all Fe moments along $-\hat{a}$. A detailed enumeration of all exchange bond contributions within the unit cell (see Supplemental Material \cite{Supplement}) yields the following total energies for a field $\mathbf{B} = B\hat{a}$:
\begin{align}
\begin{split}
E_{I}(B) &= -8\,J_{\mathrm{Mn\text{-}Fe}}\,S^{\mathrm{Mn}} S^{\mathrm{Fe}}
- 2B\left(S^{\mathrm{Mn}} - S^{\mathrm{Fe}}\right) \\
E_{II}(B) &= -12\,J_{\mathrm{Mn\text{-}Fe}}\,S^{\mathrm{Mn}} S^{\mathrm{Fe}}
+ 6\,J_{\mathrm{Mn\text{-}Mn}}\,(S^{\mathrm{Mn}})^2 \\
&+ 6\,J_{\mathrm{Fe\text{-}Fe}}\,(S^{\mathrm{Fe}})^2 - 4B\left(S^{\mathrm{Mn}} - S^{\mathrm{Fe}}\right).
\end{split}
\label{eq:EIIB}
\end{align}

At zero field, Configuration~I is the ground state provided $E_I(0) < E_{II}(0)$, i.e. $6\,J_{\mathrm{Mn\text{-}Mn}}\,(S^{\mathrm{Mn}})^2 + 6\,J_{\mathrm{Fe\text{-}Fe}}\,(S^{\mathrm{Fe}})^2 > 4\,J_{\mathrm{Mn\text{-}Fe}}\,S^{\mathrm{Mn}} S^{\mathrm{Fe}}$ (full derivation in the Supplemental Material \cite{Supplement}). Since $J_{\mathrm{Mn\text{-}Fe}} > 0$, the right-hand side is positive, so this condition can only be satisfied if $J_{\mathrm{Mn\text{-}Mn}}$ (and/or $J_{\mathrm{Fe\text{-}Fe}}$) is itself positive -- i.e., antiferromagnetic under the sign convention of Eq.~\eqref{eq:hamiltonian} -- and large enough to outweigh the Mn--Fe term. This confirms that the partially frustrated $\mathbf{k}=(1,0,0)$ ground state observed at zero field arises from this geometric frustration of the antiferromagnetic Mn--Mn coupling on the FCC sublattice.

The key difference in the Zeeman term between the two configurations reflects their distinct net magnetizations: Configuration~I carries a reduced net moment due to the partial frustration, while Configuration~II carries the full ferrimagnetic moment. Since $S^{\mathrm{Mn}} > S^{\mathrm{Fe}}$, an applied field progressively lowers the energy of Configuration~II relative to Configuration~I, driving a field-induced transition at a critical field $B_c$ defined by $E_{I}(B_c) = E_{II}(B_c)$:
\begin{equation}
\begin{split}
&-4\,J_{\mathrm{Mn\text{-}Fe}}\,S^{\mathrm{Mn}} S^{\mathrm{Fe}}
+ 6\,J_{\mathrm{Mn\text{-}Mn}}\,(S^{\mathrm{Mn}})^2
+ 6\,J_{\mathrm{Fe\text{-}Fe}}\,(S^{\mathrm{Fe}})^2\\
& = 2\,B_c\left(S^{\mathrm{Mn}} - S^{\mathrm{Fe}}\right).
\label{eq:Bc}
\end{split}
\end{equation}

Using the experimentally determined ordered moments $\mu_{\mathrm{Mn}} = 3.52\,\mu_B$ and $\mu_{\mathrm{Fe}} = 0.59\,\mu_B$, and the measured critical field $B_c = 1.3$~T, we evaluate both sides of Eq.~\eqref{eq:Bc}. With $\mu_B = 0.06$~meV.T$^{-1}$, the right-hand side yields $2\,B_c(S^{\mathrm{Mn}} - S^{\mathrm{Fe}}) \approx 0.45$~meV. The dominant contribution to the zero-field exchange energy difference arises from the Mn--Mn term, owing to the large Mn ordered moment. The small net energy difference of $\sim 0.5$~meV therefore requires a near-cancellation between the antiferromagnetic Mn--Fe coupling and the frustrated antiferromagnetic Mn--Mn interaction. Neglecting the comparatively small Fe--Fe contribution:
\begin{equation}
J_{\mathrm{Mn\text{-}Fe}} \sim
\frac{6\,(S^{\mathrm{Mn}})^2}{4\,S^{\mathrm{Mn}} S^{\mathrm{Fe}}}\,J_{\mathrm{Mn\text{-}Mn}}
\sim 8\,J_{\mathrm{Mn\text{-}Mn}}.
\end{equation}

The observation of a spin reorientation at the modest field of 1.3~T places the exchange energy scale well below 1~meV. Within a mean-field Heisenberg framework, using $T_C = 12$~K, an effective moment $\mu \approx 3\,\mu_B$, and a coordination number $z = 6$, one estimates a leading exchange constant $|J| \approx 0.1$~meV, in qualitative agreement with the energy scale set by $B_c$, thereby providing a self-consistent picture of the magnetic interactions. This analysis reveals the following hierarchy of exchange couplings:
\begin{equation}
|J_{\mathrm{Mn\text{-}Fe}}| > |J_{\mathrm{Mn\text{-}Mn}}| \gg |J_{\mathrm{Fe\text{-}Fe}}|,
\end{equation}
in which the Fe--Fe exchange plays only a marginal role. The small value of $B_c$ constitutes direct experimental evidence that MnFe sits in the vicinity of a compensation point arising from competing exchange interactions, placing the system in a regime characteristic of metamagnetic behavior.

\subsection{Dehydration-induced transition}
\label{sec:mnfedehydration}
Neutron diffraction measurements on dehydrated MnFe (Fig.~\ref{figMnFe}, Sec.~\ref{sec:dehydration_structure}) reveal that the magnetic contribution shifts to coincide with the nuclear Bragg peaks, corresponding to a propagation vector $\mathbf{k} = (0,0,0)$, analogous to the behavior observed under an applied magnetic field. However, unambiguous determination of the magnetic structure is challenging due to the presence of only a single prominent magnetic peak. This structural and magnetic modification is corroborated by magnetization measurements, where the dehydrated sample exhibits a clear bifurcation between ZFC and FC curves, indicating a change in the magnetic ground state.

Water removal similarly affects the magnetism of the other family members: CW analysis shows no sign change of $\Theta$ upon dehydration for any compound, indicating that the dominant magnetic interactions between the A- and B-site 3$d$ transition-metal ions remain essentially unchanged, but the frustration index $|\Theta|/T_C$ of CoFe increases markedly from $\sim$1.4 (hydrated) to $\sim$7.5 (dehydrated), signalling enhanced magnetic frustration upon water removal (full quantitative comparison for MnFe, CoFe and NiFe in Supplementary Material \cite{Supplement}). This confirms that water content is central to shaping the magnetic ground state throughout this family, in the same way that it governs the structural response discussed in Sec.~\ref{sec:dehydration_structure}.

\subsection{Resolution of the XMCD ambiguity}
This field-induced transition to the $I_4/mm'm'$ magnetic space group also accounts for the anomalous Fe $K$-edge XMCD response previously reported for MnFe compounds in comparison with other $A$Fe members of the PBA family: the similarity between the Mn $K$-edge XMCD behavior of MnFe and MnCr PBA compounds~\cite{n2022toward} is hereby resolved, as both systems share the same high-field magnetic symmetry. Indeed, significant effort has been devoted to investigating PBA and related molecular compounds using transition metal K-edge XMCD, where explicit relationships between XMCD intensity and total spin were established, providing crucial insights into exchange interactions and the orientation of local magnetic moments \cite{n2022toward, n2022interplay, n2024interplay}. For the AFe compounds, the magnetic structures were initially inferred from Fe $K$-edge XMCD data. However, the origin of certain distinct behaviors remained unclear. Neutron diffraction resolves this ambiguity by showing that, for MnFe, the magnetic ground state undergoes a transition around 1.3~T, whereas XMCD measurements were conducted at 1.3~T. Consequently, the XMCD did not probe the zero-field magnetic ground state of MnFe; it instead correctly captured the field-induced $I4/mm'm'$ state. This also explains why Fe $K$-edge XMCD shows a similar response across the whole AFe series (CoFe, NiFe, MnFe): under the applied field used in the XMCD experiments ($\sim$1.3~T, at or above $B_c$), MnFe is already driven into the same collinear $\mathbf{k}=(0,0,0)$ symmetry natively adopted at zero field by CoFe and NiFe, so the zero-field structural distinction revealed here by neutron diffraction was not observed.

\section{Conclusion}
Neutron diffraction provides critical insights into the magnetic moments of individual atoms that cannot be obtained from bulk magnetization measurements. Using this technique, we have investigated the structural and magnetic properties of AFe-based PBA compounds as a function of temperature and applied magnetic field.

At room temperature, all bimetallic PBAs of the Fm$\overline{3}$m family exhibit a broad low-angle diffuse feature in their neutron diffraction patterns, which we assign to vacancy-related correlations through a comparative study with Rb$_2$NiFe. High-temperature diffraction measurements reveal a crossover from positive to negative thermal expansion in CoFe and MnFe upon dehydration, while NiFe remains structurally stable.

In the magnetically ordered state, a field-induced transition from antiferromagnetic ($\mathbf{k} = (1,0,0)$) to ferrimagnetic ($\mathbf{k} = (0,0,0)$) ordering was observed in MnFe at a critical field of only 1.3~T. This spin reorientation is driven by competing nearest-neighbor Mn--Fe and next-nearest-neighbor Mn--Mn exchange interactions, with a hierarchy $|J_{\mathrm{Mn\text{-}Fe}}| > |J_{\mathrm{Mn\text{-}Mn}}| \gg |J_{\mathrm{Fe\text{-}Fe}}|$, placing MnFe in the vicinity of a magnetic compensation point. The field-induced high-symmetry state ($I4/mm'm'$) is also recovered upon dehydration, and explains the previously anomalous Fe $K$-edge XMCD response of MnFe at fields above 1.3~T.

Beyond resolving this specific XMCD ambiguity, the present neutron diffraction study provides several pieces of information that are not directly accessible to $K$-edge XMCD spectroscopy. First, it yields absolute, site-resolved ordered magnetic moments (e.g., $\mu_{\mathrm{Mn}} = 3.52\,\mu_B$, $\mu_{\mathrm{Fe}} = -0.59\,\mu_B$ in MnFe) determined directly from the magnetic Bragg intensities, internally normalized by the nuclear structure factor. Second, the field-dependent measurements reveal, for the first time, the existence of a genuine field-induced magnetic phase transition in MnFe at $B_c = 1.3$~T. Third, the analysis of the two competing magnetic configurations provides a first, admittedly approximate, estimate of the exchange energy scale ($|J| \sim 0.1$~meV) and of the resulting hierarchy of exchange pathways ($|J_{\mathrm{Mn\text{-}Fe}}| > |J_{\mathrm{Mn\text{-}Mn}}| \gg |J_{\mathrm{Fe\text{-}Fe}}|$), neither of which is accessible from a spectroscopic probe of local electronic structure such as XMCD. Finally, the partially frustrated $\mathbf{k} = (1,0,0)$ magnetic structure identified here provides a microscopic, structural origin for the pronounced non-saturation of the MnFe $M$--$H$ curve up to 7~T (Fig.~\ref{figMsPBA}) -- a phenomenon already visible in bulk magnetometry, but whose structural origin only neutron diffraction can reveal. Taken together, these results are therefore genuinely complementary to the existing studies of this compound family.

For CoFe and NiFe, only minor changes in peak intensities were detected below T$_C$, which limits determination of their magnetic structures. In contrast, MnCr exhibited a robust magnetic signal enabling a reliable magnetic structure solution. Temperature-dependent neutron diffraction up to 450~K reveals that dehydration is facilitated in compounds with larger lattice parameters, explaining the enhanced structural flexibility of CoFe and MnFe and their stronger response to water removal compared to NiFe. These results highlight the interplay between lattice size, dehydration, and magnetic stability in PBA compounds, with implications for both fundamental understanding and potential battery applications.

\section*{Acknowledgments}
Authors gratefully acknowledges funding from SAMBA project from APICONE. Experiments at ILL were sponsored by the French Neutron Federation (2FDN), with data references doi:10.5291/ILL-DATA.EASY-1598, doi:10.5291/ILL-DATA.5-21-1205 and doi:10.5291/ILL-DATA.5-31-3090.

\bibliographystyle{apsrev}
\bibliography{PRB_Draft_1}

@article{n2022toward,
  title={Toward Quantitative Magnetic Information from Transition Metal K-Edge XMCD of Prussian Blue Analogs},
  author={N’Diaye, Adama and Bordage, Am{\'e}lie and Nataf, Lucie and Baudelet, Fran{\c{c}}ois and Rivi{\`e}re, Eric and Bleuzen, Anne},
  journal={Inorganic Chemistry},
  volume={61},
  number={16},
  pages={6326--6336},
  year={2022},
  publisher={ACS Publications}
}

@article{peng2022prussian,
  title={Prussian blue analogues for sodium-ion batteries: past, present, and future},
  author={Peng, Jian and Zhang, Wang and Liu, Qiannan and Wang, Jiazhao and Chou, Shulei and Liu, Huakun and Dou, Shixue},
  journal={Advanced Materials},
  volume={34},
  number={15},
  pages={2108384},
  year={2022},
  publisher={Wiley Online Library}
}

@article{yi2021structure,
  title={Structure and properties of Prussian blue analogues in energy storage and conversion applications},
  author={Yi, Haocong and Qin, Runzhi and Ding, Shouxiang and Wang, Yuetao and Li, Shunning and Zhao, Qinghe and Pan, Feng},
  journal={Advanced Functional Materials},
  volume={31},
  number={6},
  pages={2006970},
  year={2021},
  publisher={Wiley Online Library}
}

@article{song2022confinement,
  title={Confinement of prussian blue analogs boxes inside conducting polymer nanotubes enables significantly enhanced catalytic performance for water treatment},
  author={Song, Na and Ren, Siyu and Zhang, Yue and Wang, Ce and Lu, Xiaofeng},
  journal={Advanced Functional Materials},
  volume={32},
  number={34},
  pages={2204751},
  year={2022},
  publisher={Wiley Online Library}
}

@article{xu2022pba,
  title={PBA-derived FeCo alloy with core-shell structure embedded in 2D N-doped ultrathin carbon sheets as a bifunctional catalyst for rechargeable Zn-air batteries},
  author={Xu, Xiaoqin and Xie, Jiahao and Liu, Bin and Wang, Rongyue and Liu, Mingyang and Zhang, Jun and Liu, Jin and Cai, Zhuang and Zou, Jinlong},
  journal={Applied Catalysis B: Environmental},
  volume={316},
  pages={121687},
  year={2022},
  publisher={Elsevier}
}

@article{jiang2021recent,
  title={Recent advances of prussian blue-based wearable biosensors for healthcare},
  author={Jiang, Yu and Yang, Yupeng and Shen, Liuxue and Ma, Junlin and Ma, Hongting and Zhu, Nan},
  journal={Analytical Chemistry},
  volume={94},
  number={1},
  pages={297--311},
  year={2021},
  publisher={ACS Publications}
}

@article{song2021prussian,
  title={Prussian blue analogs and their derived nanomaterials for electrochemical energy storage and electrocatalysis},
  author={Song, Xuezhi and Song, Shuyan and Wang, Dan and Zhang, Hongjie},
  journal={Small Methods},
  volume={5},
  number={4},
  pages={2001000},
  year={2021},
  publisher={Wiley Online Library}
}

@article{su2025ni,
  title={Ni-Co PBA nanocubes for efficient capture and incorporation of Cs and Tl ions: Mechanism and techno-economic analysis},
  author={Su, Minhua and Han, Weixing and Wang, Shuwen and Lai, Yanrong and Chen, Miaoling and Chen, Diyun and Song, Gang and Xu, Junhua and Tang, Jinfeng},
  journal={Journal of Environmental Chemical Engineering},
  volume={13},
  number={3},
  pages={116356},
  year={2025},
  publisher={Elsevier}
}

@article{wang2022effect,
  title={Effect of eliminating water in Prussian blue cathode for sodium-ion batteries},
  author={Wang, Wanlin and Gang, Yong and Peng, Jian and Hu, Zhe and Yan, Zichao and Lai, Weihong and Zhu, Yanfang and Appadoo, Dominique and Ye, Mao and Cao, Yuliang and others},
  journal={Advanced Functional Materials},
  volume={32},
  number={25},
  pages={2111727},
  year={2022},
  publisher={Wiley Online Library}
}

@article{yao2024achievements,
  title={Achievements in the Photomagnetic Effect of Cobalt-Iron Prussian Blue Analogues},
  author={Yao, Kangkang and Dong, Chengwei and Cao, Kaiyan and Zhang, Yin and Yang, Sen},
  journal={Advanced Functional Materials},
  volume={34},
  number={30},
  pages={2313938},
  year={2024},
  publisher={Wiley Online Library}
}

@article{fornasieri2018magnetism,
  title={Magnetism and photomagnetism of prussian blue analogue nanoparticles embedded in porous metal oxide ordered nanostructures},
  author={Fornasieri, Giulia and Bordage, Am{\'e}lie and Bleuzen, Anne},
  journal={European Journal of Inorganic Chemistry},
  volume={2018},
  number={3-4},
  pages={259--271},
  year={2018},
  publisher={Wiley Online Library}
}

@article{perez2015symmetry,
  title={Symmetry-based computational tools for magnetic crystallography},
  author={Perez-Mato, JM and Gallego, SV and Tasci, ES and Elcoro, LU{\.I}S and de la Flor, Gemma and Aroyo, MI},
  journal={Annual Review of Materials Research},
  volume={45},
  number={1},
  pages={217--248},
  year={2015},
  publisher={Annual Reviews}
}

@article{n2022interplay,
  title={Interplay between Transition-Metal K-edge XMCD and Magnetism in Prussian Blue Analogs},
  author={N’diaye, Adama and Bordage, Am{\'e}lie and Nataf, Lucie and Baudelet, Fran{\c{c}}ois and Rivi{\`e}re, Eric and Bleuzen, Anne},
  journal={ACS omega},
  volume={7},
  number={41},
  pages={36366--36378},
  year={2022},
  publisher={ACS Publications}
}

@article{n2024interplay,
  title={Interplay between transition-metal K-edge XMCD, slight structural distortions and magnetism in a series of trimetallic (Co x Ni (1- x)) 4 [Fe (CN) 6] 3/8 Prussian blue analogues},
  author={N’Diaye, Adama and Bordage, Am{\'e}lie and Nataf, Lucie and Baudelet, Fran{\c{c}}ois and Rivi{\`e}re, Eric and Bleuzen, Anne},
  journal={Physical Chemistry Chemical Physics},
  volume={26},
  number={21},
  pages={15576--15586},
  year={2024},
  publisher={Royal Society of Chemistry}
}

@article{gallego2012magnetic,
  title={Magnetic symmetry in the Bilbao Crystallographic Server: a computer program to provide systematic absences of magnetic neutron diffraction},
  author={Gallego, Samuel V and Tasci, Emre S and Flor, G and Perez-Mato, J Manuel and Aroyo, Mois I},
  journal={Applied Crystallography},
  volume={45},
  number={6},
  pages={1236--1247},
  year={2012},
  publisher={International Union of Crystallography}
}

@article{adak2011thermal,
  title={Thermal expansion in 3d-metal Prussian Blue Analogs—A survey study},
  author={Adak, Sourav and Daemen, Luke L and Hartl, Monika and Williams, Darrick and Summerhill, Jennifer and Nakotte, Heinz},
  journal={Journal of solid state chemistry},
  volume={184},
  number={11},
  pages={2854--2861},
  year={2011},
  publisher={Elsevier}
}

@article{franz2004crystalline,
  title={Crystalline, mixed-valence manganese analogue of Prussian blue: Magnetic, spectroscopic, X-ray and neutron diffraction studies},
  author={Franz, Patrick and Ambrus, Christina and Hauser, Andreas and Chernyshov, Dmitry and Hostettler, Marc and Hauser, J{\"u}rg and Keller, Lukas and Kr{\"a}mer, Karl and Stoeckli-Evans, Helen and Pattison, Philip and others},
  journal={Journal of the American Chemical Society},
  volume={126},
  number={50},
  pages={16472--16477},
  year={2004},
  publisher={ACS Publications}
}

@article{matsuda2009universal,
  title={Universal thermal response of the Prussian blue lattice},
  author={Matsuda, T and Kim, JE and Ohoyama, K and Moritomo, Y},
  journal={Physical Review B—Condensed Matter and Materials Physics},
  volume={79},
  number={17},
  pages={172302},
  year={2009},
  publisher={APS}
}

@article{margadonna2004zero,
  title={Zero thermal expansion in a Prussian blue analogue},
  author={Margadonna, Serena and Prassides, Kosmas and Fitch, Andrew N},
  journal={Journal of the American Chemical Society},
  volume={126},
  number={47},
  pages={15390--15391},
  year={2004},
  publisher={ACS Publications}
}

@article{Rodriguez1993,
  author = {Rodríguez-Carvajal, J.},
  title = {Recent advances in magnetic structure determination by neutron powder diffraction},
  journal = {Physica B: Condensed Matter},
  year = {1993},
  volume = {192},
  pages = {55--69}
}

@article{tokoro2011novel,
  title={Novel magnetic functionalities of Prussian blue analogs},
  author={Tokoro, Hiroko and Ohkoshi, Shin-ichi},
  journal={Dalton Transactions},
  volume={40},
  number={26},
  pages={6825--6833},
  year={2011},
  publisher={Royal Society of Chemistry}
}

@article{zhang2022lithiated,
  title={Lithiated Prussian blue analogues as positive electrode active materials for stable non-aqueous lithium-ion batteries},
  author={Zhang, Ziheng and Avdeev, Maxim and Chen, Huaican and Yin, Wen and Kan, Wang Hay and He, Guang},
  journal={Nature Communications},
  volume={13},
  number={1},
  pages={7790},
  year={2022},
  publisher={Nature Publishing Group UK London}
}

@article{lu2006tuning,
  title={Tuning the Magnetic Behavior via Dehydration/Hydration Treatment of a New Ferrimagnet with the Composition of K0. 2Mn1. 4Cr (CN) 6⊙ 6H2O},
  author={L{\"u}, Zhengliang and Wang, Xinyi and Liu, Zhiliang and Liao, Fuhui and Gao, Song and Xiong, Rengen and Ma, Hongwei and Zhang, Deqing and Zhu, Daoben},
  journal={Inorganic Chemistry},
  volume={45},
  number={3},
  pages={999--1004},
  year={2006},
  publisher={ACS Publications}
}

@article{temleitner2015neutron,
  title={Neutron diffraction of hydrogenous materials: Measuring incoherent and coherent intensities separately},
  author={Temleitner, L{\'a}szl{\'o} and Stunault, Anne and Cuello, Gabriel J and Pusztai, L{\'a}szl{\'o}},
  journal={Physical Review B},
  volume={92},
  number={1},
  pages={014201},
  year={2015},
  publisher={APS}
}

@article{kaye2005hydrogen,
  title={Hydrogen Storage in the Dehydrated Prussian Blue Analogues M3 [Co (CN) 6] 2 (M= Mn, Fe, Co, Ni, Cu, Zn)},
  author={Kaye, Steven S and Long, Jeffrey R},
  journal={Journal of the American Chemical Society},
  volume={127},
  number={18},
  pages={6506--6507},
  year={2005},
  publisher={ACS Publications}
}

@article{goodwin2005guest,
  title={Guest-Dependent Negative Thermal Expansion in Nanoporous Prussian Blue Analogues MIIPtIV (CN) 6⊙ x $\{$H2O$\}$(0≤ x≤ 2; M= Zn, Cd)},
  author={Goodwin, Andrew L and Chapman, Karena W and Kepert, Cameron J},
  journal={Journal of the American Chemical Society},
  volume={127},
  number={51},
  pages={17980--17981},
  year={2005},
  publisher={ACS Publications}
}

@article{chapman2006compositional,
  title={Compositional dependence of negative thermal expansion in the prussian blue analogues MIIPtIV (CN) 6 (M= Mn, Fe, Co, Ni, Cu, Zn, Cd)},
  author={Chapman, Karena W and Chupas, Peter J and Kepert, Cameron J},
  journal={Journal of the American Chemical Society},
  volume={128},
  number={21},
  pages={7009--7014},
  year={2006},
  publisher={ACS Publications}
}

@article{auckett2018continuous,
  title={Continuous negative-to-positive tuning of thermal expansion achieved by controlled gas sorption in porous coordination frameworks},
  author={Auckett, Josie E and Barkhordarian, Arnold A and Ogilvie, Stephen H and Duyker, Samuel G and Chevreau, Hubert and Peterson, Vanessa K and Kepert, Cameron J},
  journal={Nature Communications},
  volume={9},
  number={1},
  pages={4873},
  year={2018},
  publisher={Nature Publishing Group UK London}
}

@misc{Supplement,
author = Dhami,
title = {See Supplementary Information},
note = {Supplementary materials are available online. The URL will be inserted by the publisher.}
}

@article{gao2018tunable,
  title={Tunable thermal expansion from negative, zero, to positive in cubic prussian blue analogues of GaFe (CN) 6},
  author={Gao, Qilong and Shi, Naike and Sanson, Andrea and Sun, Yu and Milazzo, Ruggero and Olivi, Luca and Zhu, He and Lapidus, Saul H and Zheng, Lirong and Chen, Jun and others},
  journal={Inorganic Chemistry},
  volume={57},
  number={22},
  pages={14027--14030},
  year={2018},
  publisher={ACS Publications}
}

@article{lahiri2016understanding,
  title={Understanding temperature and magnetic-field actuated magnetization polarity reversal in the Prussian blue analogue Cu0. 73Mn0. 77 [Fe (CN) 6]. zH2O, using XMCD},
  author={Lahiri, Debdutta and Choi, Yongseong and Yusuf, SM and Kumar, Amit and Ramanan, Nitya and Chattopadhyay, Soma and Haskel, Daniel and Sharma, Surinder M},
  journal={Materials Research Express},
  volume={3},
  number={3},
  pages={036101},
  year={2016},
  publisher={IOP Publishing}
}

@article{adak2024thermal,
  title={Thermal expansion in Prussian Blue analogs M3 [Cr (CN) 6] 2. nH2O (M= Mn, Fe, Co, Ni)},
  author={Adak, Sourav and Daemen, Luke L and Hartl, Monika and Pandey, Aman Kumar and Nakotte, Heinz},
  journal={Solid State Sciences},
  volume={157},
  pages={107712},
  year={2024},
  publisher={Elsevier}
}

@article{ohkoshi2005temperature,
  title={Temperature-and photo-induced phase transition in rubidium manganese hexacyanoferrate},
  author={Ohkoshi, Shin-ichi and Tokoro, Hiroko and Hashimoto, Kazuhito},
  journal={Coordination chemistry reviews},
  volume={249},
  number={17-18},
  pages={1830--1840},
  year={2005},
  publisher={Elsevier}
}

@article{yusuf2012magnetic,
  title={Magnetic field driven transition from an antiferromagnetic ground state to a ferrimagnetic state in Rb0. 19Ba0. 3Mn1. 1 [Fe (CN) 6]{\textperiodcentered} 0.48 H2O Prussian blue analogue},
  author={Yusuf, SM and Thakur, N and Medarde, M and Keller, L},
  journal={Journal of Applied Physics},
  volume={112},
  number={9},
  year={2012},
  publisher={AIP Publishing}
}

@article{kumar2007variation,
  title={Variation of structural and magnetic properties with composition in the (Co x Ni 1- x) 1.5 [Fe (CN) 6]∙ z H 2 O series},
  author={Kumar, Amit and Yusuf, SM and Keller, L and Yakhmi, JV and Srivastava, JK and Paulose, PL},
  journal={Physical Review B—Condensed Matter and Materials Physics},
  volume={75},
  number={22},
  pages={224419},
  year={2007},
  publisher={APS}
}

@article{perezmato2024guidelines,
  author  = {Perez-Mato, J. M. and others},
  title   = {Guidelines for communicating commensurate magnetic structures. A report of the {IUCr} Commission on Magnetic Structures},
  journal = {Acta Crystallographica Section B},
  volume  = {80},
  pages   = {219--234},
  year    = {2024},
  doi     = {10.1107/S2052520624004268}
}

\end{document}


\title{Supplemental Material for : Water, vacancies, and competing exchange interactions in Prussian blue analogues: a neutron diffraction study of field- and dehydration-driven magnetic transitions}

\author{N. S. Dhami}
\email{naveen.dhami@universite-paris-saclay.fr}
\affiliation{Université Paris-Saclay, CNRS, Laboratoire de Physique des Solides, UMR-8502, 91405, Orsay, France}
\affiliation{Synchrotron SOLEIL, L’Orme des Merisiers, Saint Aubin BP 48, 91192 Gif-sur-Yvette, France}

\author{C. V. Colin}
\affiliation{Institut Néel, Université Grenoble Alpes \& CNRS, Grenoble, 38042, France}

\author{V. Nassif}
\affiliation{Institut Néel, Université Grenoble Alpes \& CNRS, Grenoble, 38042, France}

\author{O. Fabelo}
\affiliation{Institut Laue-Langevin, 38000 Grenoble, France}

\author{T. Nait}
\affiliation{ICMMO, Université Paris Saclay, CNRS, 15 rue Georges Clémenceau, 91405 Orsay, France}

\author{A. Bleuzen}
\affiliation{ICMMO, Université Paris Saclay, CNRS, 15 rue Georges Clémenceau, 91405 Orsay, France}

\author {A. Bordage}
\email{amelie.bordage@universite-paris-saclay.fr}
\affiliation{ICMMO, Université Paris Saclay, CNRS, 15 rue Georges Clémenceau, 91405 Orsay, France}

\author{V. Balédent}
\email{victor.baledent@universite-paris-saclay.fr}
\affiliation{Université Paris-Saclay, CNRS, Laboratoire de Physique des Solides, UMR-8502, 91405, Orsay, France}
\affiliation{Institut universitaire de France (IUF)}

\maketitle

\setcounter{figure}{0}
\renewcommand{\thefigure}{S\arabic{figure}}
\setcounter{table}{0}
\renewcommand{\thetable}{S\arabic{table}}

\section{Synthesis and composition}

\subsection{Synthesis protocol}
The PBAs were synthesized by a drop-by-drop addition of two precursors in aqueous solutions: a 400~mL aqueous solution of potassium hexacyanoferrate(III) K$_3$[Fe(CN)$_6$] or potassium hexacyanochromate(III) K$_3$[Cr(CN)$_6$] ($c = 2.5\times 10^{-3}$~mol.L$^{-1}$) was added to a 100~mL aqueous solution of nitrate salts A(NO$_3$)$_2$ ($A$ = Mn, Fe, Co, Ni, Cu; $c = 50\times 10^{-3}$~mol.L$^{-1}$). The precipitates were washed with distilled water and centrifuged three times at 8000~rpm, and finally allowed to dry in air at room temperature, as detailed in Ref.~\cite{n2022toward}. When the synthesis is realized in the presence of alkali cations (Rb), the alkali cation may be trapped in the interstitial sites of the structure. All syntheses were performed by the ICMMO synthesis platform. A detailed description of the sample characterization is also provided in Ref.~\cite{n2022toward}.

\subsection{Composition: EDS and TGA}

The metal stoichiometry (A:B ratio) of each compound was determined by energy-dispersive X-ray spectroscopy (EDS), and the interstitial/coordinated water content was determined by thermogravimetric analysis (TGA). Table~\ref{tab:EDS} summarizes the experimental EDS atomic percentages, compared to the values calculated from the nominal (anhydrous) chemical formula given in the main text, together with the fully hydrated formula obtained by combining the EDS-derived metal ratio with the TGA-derived water content.

\begin{table}[htbp]
\centering
\begin{tabular}{lccc}
\hline
\tiny
\textbf{Compound} & \textbf{EDS atomic \% exp.\ (calc.)} & \textbf{Anhydrous formula (EDS)} & \textbf{Hydrated formula (TGA)} \\
\hline
MnFe & Mn 59.4(59.7), Fe 40.6(40.3) & Mn$_4$[Fe(CN)$_6$]$_{2.7}$ & Mn$_4$[Fe(CN)$_6$]$_{2.7}\cdot$16.5H$_2$O \\
CoFe & Co 60.4(59.7), Fe 39.6(40.3) & Co$_4$[Fe(CN)$_6$]$_{2.7}$ & Co$_4$[Fe(CN)$_6$]$_{2.7}\cdot$16.5H$_2$O \\
NiFe & Ni 59.8(59.7), Fe 40.2(40.3) & Ni$_4$[Fe(CN)$_6$]$_{2.7}$ & Ni$_4$[Fe(CN)$_6$]$_{2.7}\cdot$16.5H$_2$O \\
Rb$_2$NiFe & Rb 21.91(21.9), Ni 41.50(41.7), Fe 36.59(35.6) & Rb$_{2.1}$Ni$_4$[Fe(CN)$_6$]$_{3.5}$ & Rb$_{2.1}$Ni$_4$[Fe(CN)$_6$]$_{3.5}\cdot$14H$_2$O) \\
MnCr & K 1.8(1.5), Mn 57.9(58.0), Cr 40.3(40.6) & K$_{0.1}$Mn$_4$[Cr(CN)$_6$]$_{2.8}$ & K$_{0.1}$Mn$_4$[Cr(CN)$_6$]$_{2.8}\cdot$16.5H$_2$O \\
\hline
\end{tabular}
\caption{EDS-determined metal stoichiometry and TGA-determined water content for the PBA compounds of this study. The trace K content in MnCr likely originates from the K$_3$[Cr(CN)$_6$] precursor used in the synthesis.}
\label{tab:EDS}
\end{table}

\begin{figure}[!htb]
    \centering
    \includegraphics[width=0.6\textwidth]{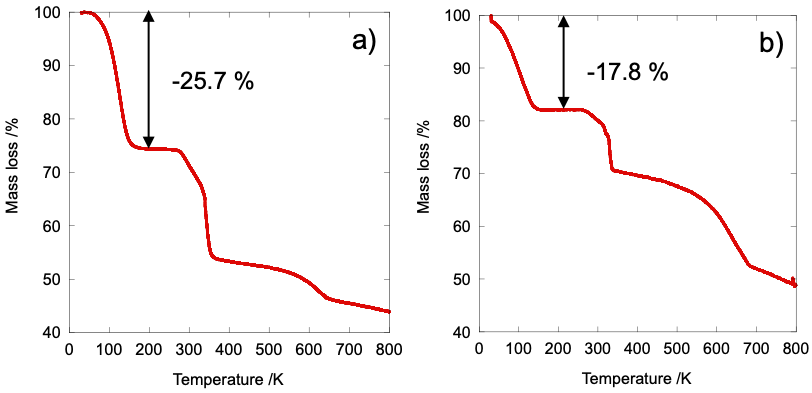}
    \caption{Representative thernogravimetric curve of a) an alkali cation-free PBAs, and b) a PBA containing two Rb+ cations per unit cell. }
    \label{figTGA}
\end{figure}

\newpage
\section{Magnetic symmetry analysis}
This section details the group-theoretical (MAXMAGN) enumeration of magnetic space groups compatible with the two propagation vectors observed in this study, $\mathbf{k} = (1,0,0)$ (MnFe, zero field) and $\mathbf{k} = (0,0,0)$ (all other compounds at zero field, and MnFe above $B_c$ or upon dehydration), summarized in the main text. Magnetic space groups were enumerated from the parent structure Fm$\overline{3}$m (\#225) using MAXMAGN~\cite{perez2015symmetry, gallego2012magnetic}, and candidate solutions were subsequently tested by Rietveld refinement of the difference data (well below $T_C$ minus well above $T_C$) using the FullProf suite~\cite{Rodriguez1993}.

\subsection{Propagation vector $\mathbf{k} = (1,0,0)$}
For $\mathbf{k} = (1,0,0)$, twelve magnetic space groups are compatible with the parent structure. Eight of these are tetragonal, of which only $P_I4/mnc$ (\#128.410) allows a non-zero ordered magnetic moment on the Mn and Fe sites; the remaining four are orthorhombic, of which two, $C_Amca$ (\#64.480) and $P_Innm$ (\#58.404), are compatible with a non-zero moment. For $P_I4/mnc$ and $C_Amca$, symmetry constrains the magnetic moment to lie along one of the cubic crystallographic axes, whereas for $P_Innm$ the moment is instead required to lie along a face diagonal of the cube. Among these three candidate magnetic space groups, Rietveld refinement of the MnFe zero-field difference data shows that only $P_I4/mnc$ provides a satisfactory fit, demonstrating that the ordered Mn and Fe moments are constrained along a cubic crystallographic axis rather than along a face diagonal.

\subsection{Propagation vector $\mathbf{k} = (0,0,0)$}
For $\mathbf{k} = 0$, the same parent group Fm$\overline{3}$m again admits twelve magnetic space groups: four cubic groups, none of which allow a non-zero ordered moment on the Mn or Fe (or Cr) sites; four tetragonal groups, of which only $I4/mm'm'$ (\#139.537) is compatible with a non-zero moment; and two orthorhombic groups, of which only $Im'm'm$ (\#71.536) is compatible with a non-zero moment. In $I4/mm'm'$, symmetry forces the magnetic moment to lie strictly along a cubic crystallographic axis, excluding both the body diagonal and the face diagonal of the cube, whereas in $Im'm'm$ the moment is instead required to lie along a face diagonal. For MnCr (zero field) and for MnFe above $B_c$ or upon dehydration, the $I4/mm'm'$ solution provides the better agreement with the data among these two candidate magnetic space groups, showing that the ordered moments are likewise constrained along a cubic crystallographic axis in all these cases.

\newpage
\section{Neutron diffraction}
\label{sec-neutron}

The neutron diffraction data was reduced with the help of lamp software. Rietveld refinement was performed using FullProf software. The refinement graphs and tables with refinement parameters for selected compounds are presented below.

\subsection{NiFe}

\begin{figure}[!htb]
    \centering
    \includegraphics[width=1.05\textwidth]{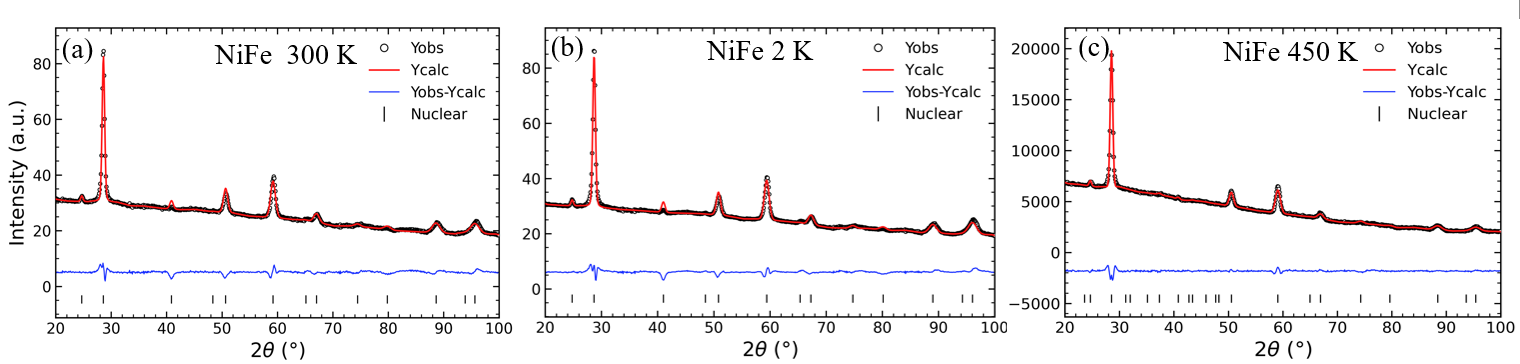}
    \caption{(a) Rietveld refinement of NiFe PBA at 300 K, 2 K and 450 K.}
    \label{figCoFe_NiFe}
\end{figure}

\begin{table}[htbp]
\centering

\label{tab:rietveld}
\begin{tabular}{lccc}
\hline
\textbf{Parameter} & \textbf{2 K} & \textbf{300 K} & \textbf{450 K} \\
\hline
Radiation, $\lambda$ (\AA)          & \multicolumn{3}{c}{Neutron, 2.5240} \\
Diffractometer                      & \multicolumn{3}{c}{D1B} \\
$2\theta$ range (\textdegree)       & \multicolumn{3}{c}{20--100} \\
Step size (\textdegree)             & \multicolumn{3}{c}{0.100078} \\
Space group                         & \multicolumn{3}{c}{$Fm\bar{3}m$ (No.\ 225)} \\
\hline
$a$ (\AA)                           & 10.187(4)     & 10.22(1)      & 10.24(1) \\
$V$ (\AA$^3$)                       & 1057.2(12)    & 1067.5(31)    & 1073.7(31) \\
Density (g/cm$^3$)                  & 1.724         & 1.707         & 1.695 \\
\hline

$R_p$ (\%)                          & 27.0          & 31.1          & 45.0 \\
$R_{wp}$ (\%)                       & 16.9          & 18.0          & 25.4 \\

$\chi^2$                            & 0.00949       & 0.00982       & 4.38 \\
GoF                                  & 0.10          & 0.10          & 2.1 \\
Bragg $R$-factor (\%)               & 13.8          & 11.5          & 30.1 \\
$R_F$-factor (\%)                   & 10.8          & 7.43          & 22.4 \\
\hline
\end{tabular}
\caption{Rietveld refinement parameters for NiFe from neutron powder diffraction data at 2, 300, and 450~K.}
\end{table}

\begin{table}[htbp]
\centering
\caption{Atomic coordinates, isotropic displacement parameters, and site occupancies at 2, 300, and 450~K for NiFe.}
\label{tab:atomic-positions}
\begin{tabular}{lccccccccc}
\hline
\textbf{Atom} & \textbf{Wyckoff} & \multicolumn{3}{c}{$x$} & $y$ & $z$ & \textbf{$U_{iso}$ (\AA$^2$)} & \textbf{Site occ.}\footnotemark[3] & \textbf{Occupancy (\%)}\footnotemark[4] \\
              &                   & 2 K & 300 K & 450 K &     &     & & & \\
\hline
Ni1 & $4b$  & \multicolumn{3}{c}{1/2}     & 0    & 0    & 0.0205  & 0.02083 & 100.00 \\
Fe1 & $4a$  & \multicolumn{3}{c}{0}       & 0    & 0    & 0.0177  & 0.01479 & 71.00  \\
C1  & $24e$ & 0.1840 & 0.1815 & 0.1815    & 0    & 0    & 0.0461  & 0.09035 & 72.29  \\
N1  & $24e$ & 0.3191 & 0.3058 & 0.3058    & 0    & 0    & 0.0512  & 0.09035 & 72.29  \\
O1  & $24e$ & 0.2852 & 0.3305 & 0.3305    & 0    & 0    & 0.1570 & 0.04125 & 33.01  \\
O2  & $32f$ & 0.0630 & 0.0800 & 0.0800    & $=x$ & $=x$ & 0.0583 & 0.02000 & 12.00  \\
O3  & $8c$  & \multicolumn{3}{c}{1/4}     & 1/4  & 1/4  & 0.1583 & 0.02083 & 50.00  \\
\hline
\end{tabular}
\footnotetext[2]{$U_{iso}$, and occupancies were refined for 300K refinement only; the scale factor, lattice parameter $a$, atomic coordinates, and profile background/shape term were refined for all temperatures. }
\footnotetext[3]{FullProf site-occupation parameter (multiplicity-weighted; not the fractional site occupancy directly).}
\footnotetext[4]{Normalized site occupation, i.e.\ fraction of the Wyckoff site occupied.}
\end{table}

\newpage
\subsection{MnFe}

\begin{figure}[!htb]
    \centering
    \includegraphics[width=1.05\textwidth]{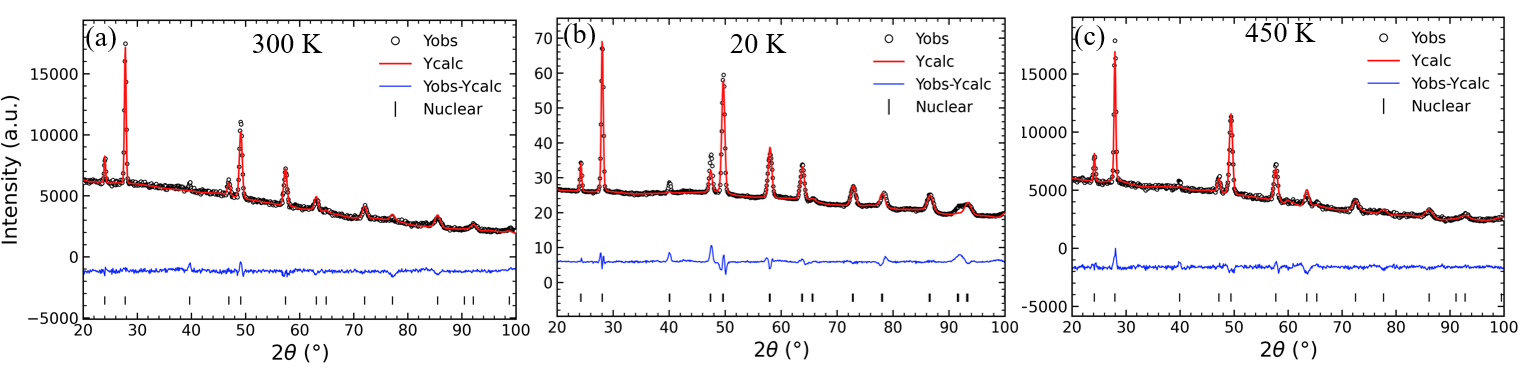}
    \caption{(a) Rietveld refinement of MnFe PBA at 300 K, 20 K and 450 K.}
    \label{figCoFe_NiFe2}
\end{figure}

\begin{table}[htbp]
\centering
\label{tab:rietveld2}
\begin{tabular}{lccc}
\hline
\textbf{Parameter} & \textbf{20 K} & \textbf{300 K} & \textbf{450 K} \\
\hline
Radiation, $\lambda$ (\AA)          & \multicolumn{3}{c}{Neutron, 2.5240} \\
Diffractometer                      & \multicolumn{3}{c}{D1B} \\
$2\theta$ range (\textdegree)       & 20--100       & 20--100       & 18--100 \\
Step size (\textdegree)             & \multicolumn{3}{c}{0.100078} \\
Space group                         & \multicolumn{3}{c}{$Fm\bar{3}m$ (No.\ 225)} \\
\hline
$a$ (\AA)                           & 10.424(2)     & 10.4980(3)    & 10.46(1) \\
$V$ (\AA$^3$)                       & 1132.6(6)     & 1156.96(6)    & 1144.4(3) \\
Density (g/cm$^3$)                  & 1.445         & 1.414         & 1.429 \\
\hline
$R_p$ (\%)                          & 23.6          & 42.8          & 40.5 \\
$R_{wp}$ (\%)                       & 18.6          & 24.7          & 22.4 \\
$R_{exp}$ (\%)                      &               & 11.6          & \\
$\chi^2$                            & 0.01327       & 4.563         & 4.639 \\
GoF                                  & 0.11          & 2.1           & 2.1 \\
Bragg $R$-factor (\%)               & 12.1          & 12.3          & 6.71 \\
$R_F$-factor (\%)                   & 8.06          & 10.3          & 4.12 \\
\hline
\end{tabular}
\caption{Rietveld refinement parameters for MnFe from neutron powder diffraction data at 20, 300, and 450~K.}
\end{table}

\begin{table}[htbp]
\centering
\caption{Atomic coordinates, isotropic displacement parameters, and site occupancies at 20, 300, and 450~K for MnFe.}
\label{tab:atomic-positions2}
\begin{tabular}{lccccccccc}
\hline
\textbf{Atom} & \textbf{Wyckoff} & \multicolumn{3}{c}{$x$} & $y$ & $z$ & \textbf{$U_{iso}$ (\AA$^2$)}\footnotemark[2] & \textbf{Site occ.}\footnotemark[3] & \textbf{Occupancy (\%)}\footnotemark[4] \\
              &                   & 20 K & 300 K & 450 K &     &     & & & \\
\hline
Mn1 & $4b$  & \multicolumn{3}{c}{1/2}     & 0    & 0    & 0.0253 & 0.02482 & 100.00 \\
Fe1 & $4a$  & \multicolumn{3}{c}{0}       & 0    & 0    & 0.0253 & 0.01686 & 67.93  \\
C1  & $24e$ & 0.1729 & 0.1683 & 0.1754    & 0    & 0    & 0.1559 & 0.10012 & 67.23  \\
N1  & $24e$ & 0.2996 & 0.2899 & 0.2942    & 0    & 0    & 0.0238 & 0.10012 & 67.23  \\
O1  & $24e$ & 0.3136 & 0.3826 & 0.3229    & 0    & 0    & 0.2532 & 0.03829 & 25.71  \\
O2  & $32f$ & 0.1134 & 0.0935 & 0.1213    & $=x$ & $=x$ & 0.0869 & 0.01898 & 9.56   \\
O3  & $8c$  & \multicolumn{3}{c}{1/4}     & 1/4  & 1/4  & 0.0753 & 0.01791 & 36.08  \\
\hline
\end{tabular}
\footnotetext[2]{$U_{iso}$, and occupancies were refined for 300K refinement only; the scale factor, lattice parameter $a$, atomic coordinates, and profile background/shape term were refined for all temperatures. $U_{iso} = B_{iso}/8\pi^2$.}
\footnotetext[3]{FullProf site-occupation parameter (multiplicity-weighted; not the fractional site occupancy directly).}
\footnotetext[4]{Normalized site occupation, i.e.\ fraction of the Wyckoff site occupied.}
\end{table}

\begin{table}[h]
\centering

\label{tab:MnFe_20K}
\begin{tabular}{lccc}
\hline
Parameter & 2 K (0 T) & 2 K (1.3 T) & 2 K (5T)\\
\hline
Parent space group & $Fm\bar{3}m$ & $Fm\bar{3}m$ & $Fm\bar{3}m$\\
Magnetic space group (BNS) &
$P_{I}4/mnc$ (No.~128.410) &
$I4/mm'm'$ (No.~139.537) &$I4/mm'm'$ (No.~139.537)  \\

$R_\mathrm{wp}$ (\%) & 53.4 & 57.5 & 30.8 \\
$R_\mathrm{exp}$ (\%) & 45.1 & 45.8 & 20.2\\
$\chi^2$ & 1.403 & 1.575& 2.334 \\
Bragg $R$ (\%) & 33.5 & 42.8 &6.19 \\

Mn ordered moment ($\mu_B$) & 3.52(5) & 3.30(9) & 5.59 (6) \\
Fe ordered moment ($\mu_B$) & $-0.59(7)$ & $-0.92(12)$ & $-1.20(8)$\\

$R_{\mathrm{Mag}}$ (\%) & 50.3 & 53.8 & 6.24\\
\hline
\end{tabular}
\caption{Summary of the Rietveld refinements of the magnetic structure for MnFe at 2 K in zero field, under an applied field of 1.3 T and 5 T.}
\end{table}

\newpage
\subsection{MnCr - Rietveld refinement and Magnetic structure}

\begin{table}[h]
\label{tab:rietveld3}
\centering
\begin{tabular}{lcc}
\hline
Parameter & 4 K (0 T) & 4 K (1.3 T) \\
\hline
Parent space group & $Fm\bar{3}m$ & $Fm\bar{3}m$ \\
Magnetic space group (BNS) & $I4/mm'm'$ (No.~139.537) & $I4/mm'm'$ (No.~139.537) \\
$R_\mathrm{p}$ (\%) & 86.5 & 86.7 \\
$R_\mathrm{wp}$ (\%) & 23.5 & 20.7 \\
$R_\mathrm{exp}$ (\%) & 35.8 & 26.1 \\
$\chi^2$ & 0.429 & 0.632 \\
Bragg $R$ (\%) & 14.8 & 35.4 \\
Mn ordered moment ($\mu_B$) & $4.90 (9)$ & 5.86(2) \\
Cr ordered moment ($\mu_B$) & - $2.63(14)$ & -3.17(4) \\
$R_{Mag}$ (\%) & 17.2 & 26.1 \\
\hline
\end{tabular}
\caption{Summary of the Rietveld refinements of the magnetic structure for MnCr at 4 K in zero field and under an applied field of 1.3 T.}
\end{table}

\begin{figure}[!htb]
    \centering
    \includegraphics[width=1.0\textwidth]{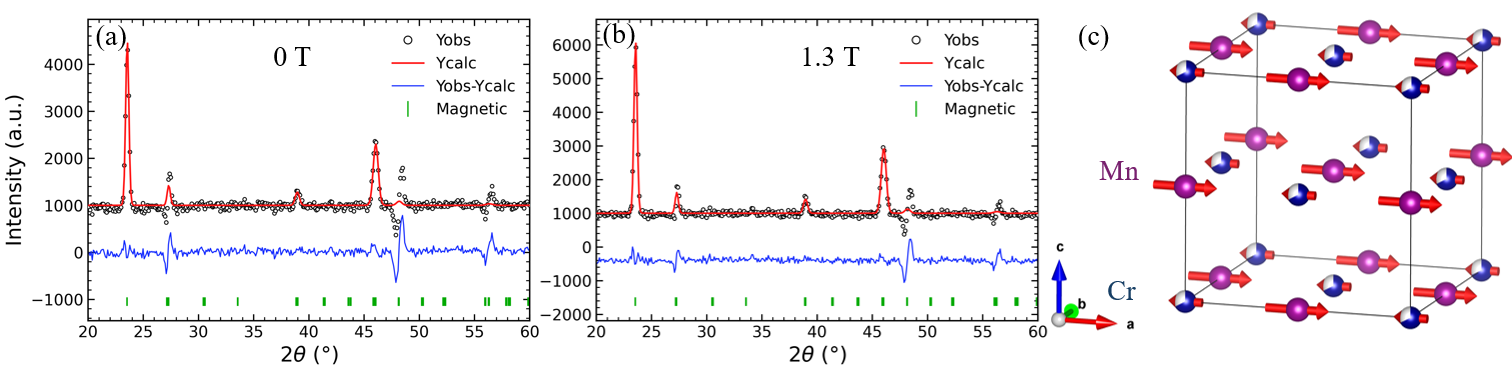}
    \caption{(a) Rietveld refinement of the difference data at 4K and 70 K, the nuclear peaks were excluded from the refinement. (b) Difference data at 1.3 T at 4 K and 70 K. (c) The magnetic structure for MnCr PBA.}
    \label{figCoFe_NiFe3}
\end{figure}

\newpage
\subsection{Magnetic structure description}
\label{sec-magcif}

Following the recent guidelines of the IUCr Commission on Magnetic Structures for communicating commensurate magnetic structures~\cite{perezmato2024guidelines}, the crystallographic description of the three magnetic structures reported in this work (MnFe at 0~T, MnFe at $B\geq1.3$~T, and MnCr at 0~T) is summarized in Tables~\ref{tab:magcif-MnFe} and~\ref{tab:magcif-MnCr} below in line with the recommendation of Perez-Mato \emph{et al.} (2024).

\begin{table}[htbp]
\centering
\footnotesize
\begin{tabular}{lccc}
\hline
 & MnFe, 2~K (0~T) & MnFe, 2~K (1.3~T) & MnFe, 2~K (5~T) \\
\hline
Parent space group & $Fm\bar{3}m$ & $Fm\bar{3}m$ & $Fm\bar{3}m$ \\
Propagation vector $\mathbf{k}$ & $(1,0,0)$ & $(0,0,0)$ & $(0,0,0)$ \\
Transformation parent$\to$magnetic basis &
$\begin{pmatrix}
0 & 0 & 1 & 0 \\
\frac{1}{2} & \frac{1}{2} & 0 & 0 \\
-\frac{1}{2} & \frac{1}{2} & 0 & 0
\end{pmatrix}$ & $\begin{pmatrix}
0 & 0 & -1 & 0 \\
\frac{1}{2} & \frac{1}{2} & 0 & 0 \\
\frac{1}{2} & -\frac{1}{2} & 0 & 0
\end{pmatrix}$ & $\begin{pmatrix}
0 & 0 & -1 & 0 \\
\frac{1}{2} & \frac{1}{2} & 0 & 0 \\
\frac{1}{2} & -\frac{1}{2} & 0 & 0
\end{pmatrix}$ \\
MSG symbol (BNS) & $P_I4/mnc$ & $I4/mm'm'$ & $I4/mm'm'$ \\
MSG number (BNS) & 128.410 & 139.537 & 139.537 \\
MSG symbol (UNI) & $P4/mnc : 1'^{I}_{0}\ [Fm\overline{3}m]$ & $I4/mm'm'$ & $I4/mm'm'$ \\
Magnetic point group (MPG) & $4/mmm$ & $4/mm'm'$ & $4/mm'm'$ \\
Mn moment components ($\mu_B$) & $(0,0,3.52(5))$ &  $(0,0,3.30(9))$ & $(0,0,5.59(6))$ \\
Fe moment components ($\mu_B$) & $(0,0,0.59(7))$ & $(0,0,0.92(12))$ & $(0,0,1.20(8))$ \\
\hline
\end{tabular}
\caption{Crystallographic description of the MnFe magnetic structures following the guidelines of Ref.~\cite{perezmato2024guidelines}. Moment magnitudes ($m_z$) are taken from Table~\ref{tab:MnFe_20K}.}
\label{tab:magcif-MnFe}
\end{table}

\begin{table}[htbp]
\centering
\footnotesize
\begin{tabular}{lcc}
\hline
 & MnCr, 4~K (0~T) & MnCr, 4~K (1.3~T) \\
\hline
Parent space group & \multicolumn{2}{c}{$Fm\bar{3}m$ (No.~225)} \\
Propagation vector $\mathbf{k}$ & $(0,0,0)$ & $(0,0,0)$ \\
Transformation parent$\to$magnetic basis & $\begin{pmatrix}
0 & 0 & -1 & 0 \\
\frac{1}{2} & \frac{1}{2} & 0 & 0 \\
\frac{1}{2} & -\frac{1}{2} & 0 & 0
\end{pmatrix}$ & $\begin{pmatrix}
0 & 0 & -1 & 0 \\
\frac{1}{2} & \frac{1}{2} & 0 & 0 \\
\frac{1}{2} & -\frac{1}{2} & 0 & 0
\end{pmatrix}$ \\
MSG symbol (BNS) & $I4/mm'm'$ & $I4/mm'm'$ \\
MSG number (BNS) & 139.537 & 139.537 \\
MSG symbol (UNI) & $I4/mm'm'$ & $I4/mm'm'$ \\
Magnetic point group (MPG) & $4/mm'm'$ & $4/mm'm'$ \\
Mn moment components ($\mu_B$) & $(0,0,4.90(9))$ &  $(0,0,5.86(2))$ \\
Cr moment components ($\mu_B$) & $(0,0,-2.63(14))$ &  $(0,0,-3.17(4))$ \\
\\
\hline
\end{tabular}
\caption{Crystallographic description of the MnCr magnetic structure following the guidelines of Ref.~\cite{perezmato2024guidelines}. Moment magnitudes ($m_z$) are taken from Table~\ref{tab:rietveld3}.}
\label{tab:magcif-MnCr}
\end{table}

Positions of the non-magnetic atoms (C, N, O framework and vacancy/water-related sites) in the same magnetic-cell setting are as given in Tables~\ref{tab:atomic-positions} and~\ref{tab:atomic-positions2} above; per Ref.~\cite{perezmato2024guidelines}, these should ideally be re-expressed directly in the magnetic (rather than purely nuclear/parent) cell setting once the transformation matrices above are confirmed, rather than reported as a separate ``two-phase'' nuclear/magnetic description.

\newpage
\section{Magnetization}

The magnetic susceptibility of all compounds was analyzed in the paramagnetic regime using the Curie--Weiss (CW) law
\[ \chi = \frac{C}{T - \theta}, \]
where $C$ is the Curie constant and $\theta$ is the Weiss temperature. The effective magnetic moment $\mu_{\rm eff}$ was obtained from the CW fits to the high-temperature susceptibility data:
\[
\mu_{\mathrm{eff}} = \sqrt{\frac{3 k_B C}{N_A}} = \sqrt{8C},
\]
where $C$ is expressed in emu.K.mol$^{-1}$. The obtained $\mu_{\rm eff}$ values were compared with the corresponding spin-only values; the resulting electronic configurations of the metal centers and their high-spin or low-spin states are in good agreement with previously reported studies~\cite{n2022toward, n2022interplay, n2024interplay} and are further supported by the neutron diffraction results presented in the main text.

For the bimetallic compounds considered here, the measured susceptibility is the sum of the independent A- and B-sublattice contributions, $\chi = \chi_A + \chi_B$, and it is therefore only at the level of the Curie constant that additivity holds: $C_{\rm tot} = n_A C_A + n_B C_B$, with $n_A$, $n_B$ the number of A- and B-site ions per formula unit. Since $\mu_{\rm eff} \propto \sqrt{C}$, this means that
\[
\mu_{\rm eff,tot} = \sqrt{n_A\, \mu_{{\rm eff},A}^2 + n_B\, \mu_{{\rm eff},B}^2}, \qquad \mu_{{\rm eff},i} = g\sqrt{S_i(S_i+1)}.
\]
The spin-only estimates given in the main text (Table~II) use this combination rule, with $g=2$ and the following electronic states, consistent with the high-spin/low-spin assignments established from the Rietveld-refined ordered moments (Sec.~\ref{sec-neutron}, and main text): Mn$^{2+}$ ($3d^5$, high-spin, $S=5/2$), Fe$^{3+}$ ($3d^5$, low-spin, $S=1/2$), and Cr$^{3+}$ ($3d^3$, $S=3/2$, the only spin state accessible to this configuration in an octahedral ligand field, independent of field strength).

\subsection{Curie Wiess Fit for PBA compounds}

\begin{figure}[!htb]
    \centering
    \includegraphics[width=1.0\textwidth]{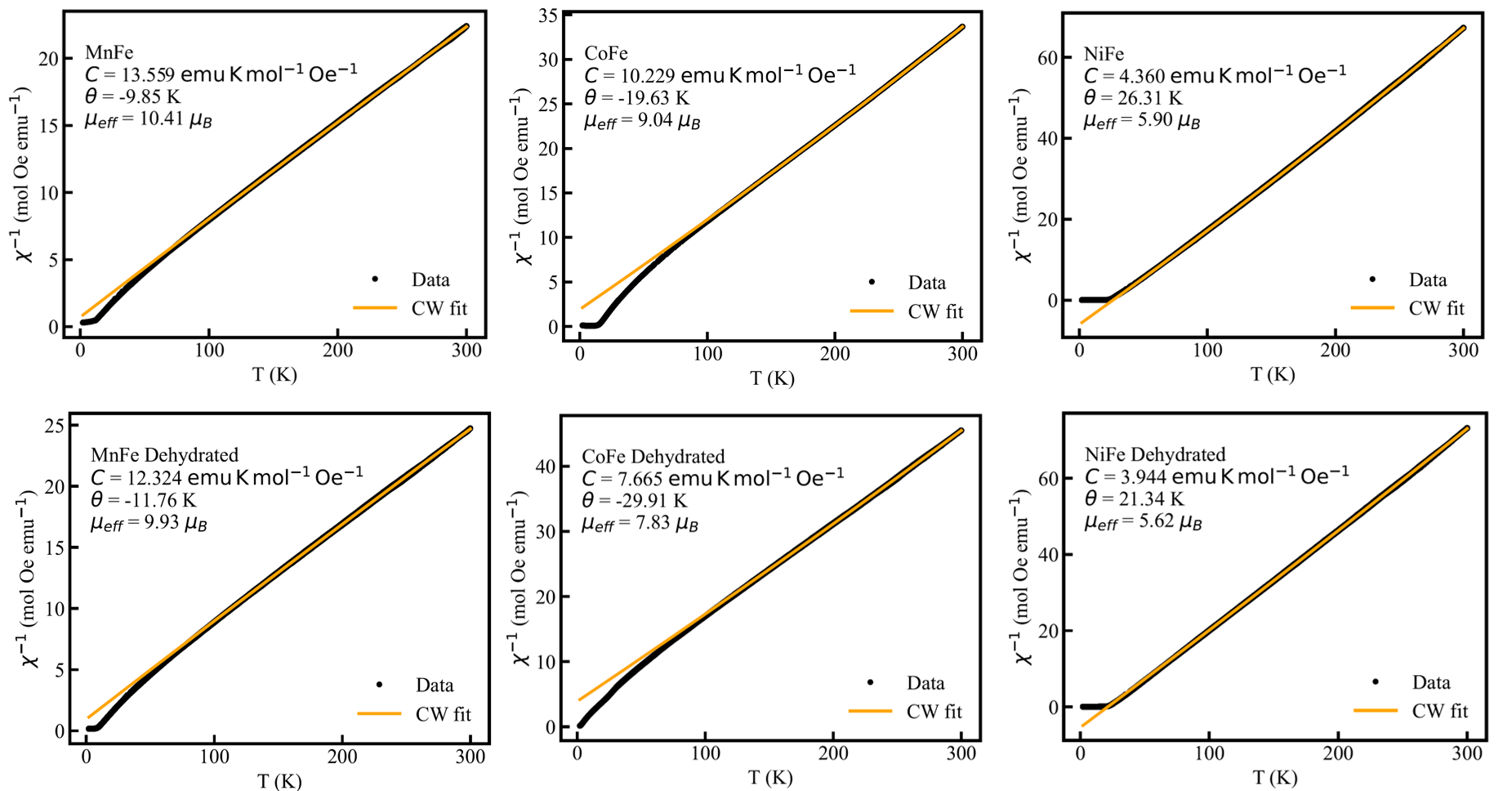}
    \caption{Effect of dehydration: The curie wiess fit before (top panels) and after (bottome panes) the dehydration for MnFe (left), CoFe (middle) and NiFe (right). The magnetization was measured under 500 Oe magnetic field. The Curie Weiss fit was performed in the temperature range 150 to 300 K.}
    \label{figCoFe_NiFe4}
\end{figure}

\begin{table*}[htb!]
\centering
\begin{tabular}{c c c}
 \hline
Composition & abbreviation & Molar mass (g/mol) \\ 

\hline
Co$_4$[Fe(CN)$_6$]$_{2.7}$ & CoFe & 808.01  \\
Mn$_4$[Fe(CN)$_6$]$_{2.7}$ & MnFe & 792.03    \\
Ni$_4$[Fe(CN)$_6$]$_{2.7}$ & NiFe & 807.05  \\
Rb$_2$Ni$_4$[Fe(CN)$_6$]$_{3.3}$ & Rb$_2$NiFe & 1105.1541   \\
Mn$_4$[Cr(CN)$_6$]$_{2.7}$ & MnCr & 781.63 \\
\hline
\end{tabular}
\caption{Compositions, respective molar mass. No water molecule (interstetial or coordinated) was taken in account for the molar mass calculation. }
\label{table1}
\end{table*}

\begin{figure}[!htb]
    \centering
    \includegraphics[width=1.05\textwidth]{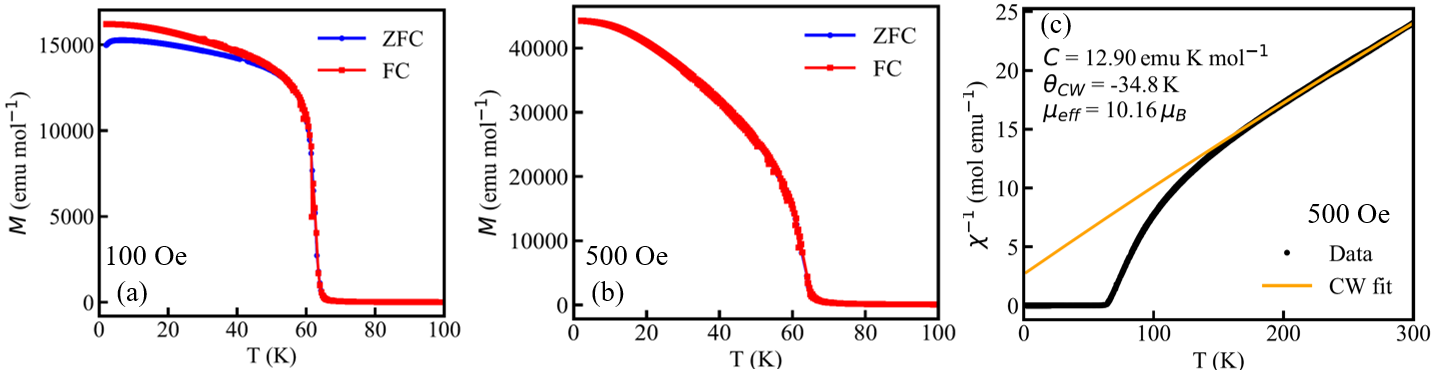}
    \caption{MnCr PBA magnetization, at 100 Oe and 500 Oe, the CW fitting at 500 Oe. The CW fitting was performed from 200 to 300 K.}
    \label{figCoFe_NiFe5}
\end{figure}

\begin{figure}[!htb]
    \centering
    \includegraphics[width=1.0\textwidth]{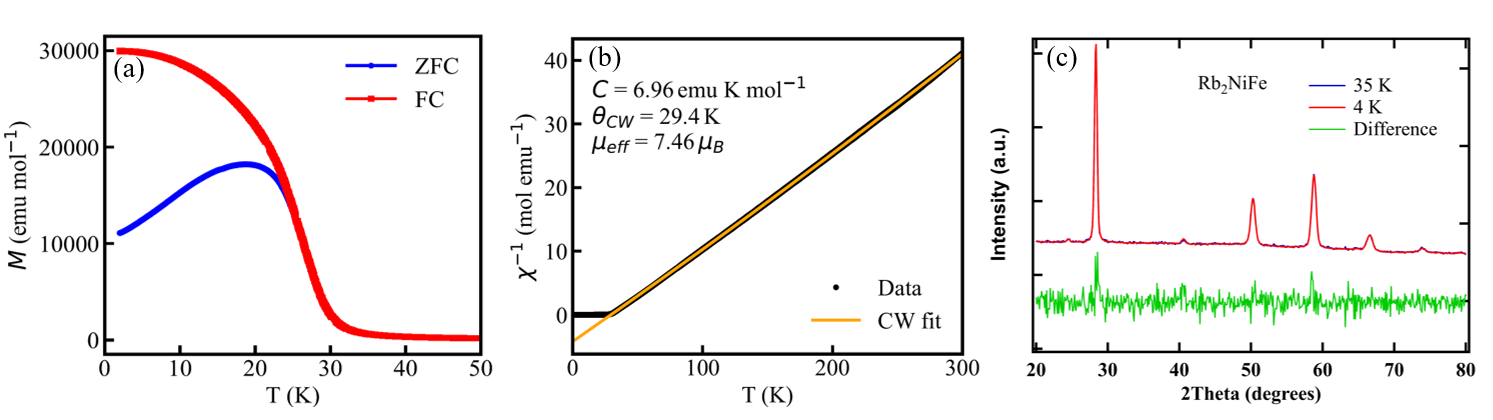}
    \caption{(a) Rb$_2$NiFe PBA magnetization under external magnetic field of 500 Oe, (b) CW fitting at 500 Oe. (c) Neutron diffraction patterns at 35 K and 4 K along with the difference, showing the expected ferromagnetic contribution on top of nuclear peaks.}
    \label{figCoFe_NiFe6}
\end{figure}

\begin{figure}[!htb]
    \centering
    \includegraphics[width=1.0\textwidth]{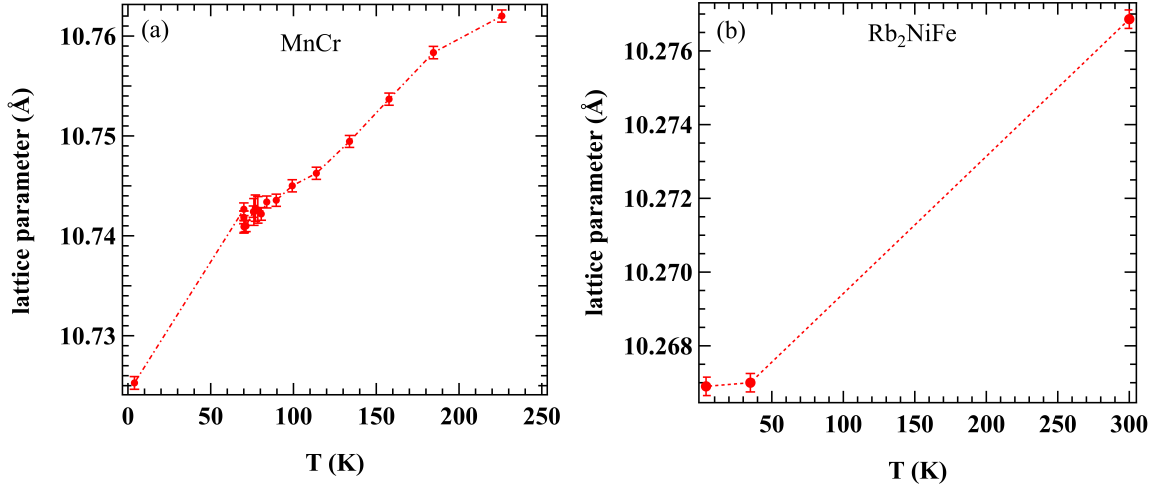}
    \caption{Temperature evolution of lattice parameter for (a) MnCr and Rb$_2$NiFe PBA compounds.}
    \label{figCoFe_NiFe7}
\end{figure}

\subsection{Effect of dehydration on magnetic properties}

Magnetization measurements were performed on the samples before and after dehydration in order to investigate the role of lattice water on their magnetic properties. The temperature dependence of the magnetization measured under zero-field-cooled (ZFC) and field-cooled (FC) conditions reveals distinctly different behaviors between the hydrated and dehydrated samples, indicating that dehydration significantly modifies the magnetic interactions within the framework.

For NiFe, the magnetization curves before and after dehydration are nearly identical, with only a slight shift in the magnetic ordering temperature from 21~K to 20~K after dehydration. In contrast, CoFe exhibits a pronounced suppression of the magnetic ordering temperature upon dehydration, shifting to significantly lower temperatures. CW analysis of the dehydrated compounds shows a slight shift in the Weiss temperature $\theta$ compared to the hydrated samples; however, no sign change of $\theta$ is observed upon dehydration, indicating that the dominant magnetic interactions between the A- and B-site 3$d$ transition-metal ions remain essentially unchanged. Notably, the frustration index $|\Theta|/T_C$ increases markedly for CoFe upon dehydration, from $\sim$1.4 in the hydrated state to $\sim$7.5 in the dehydrated state (Table~\ref{table5}), indicating that water removal substantially enhances magnetic frustration in this compound.

The correlation between crystal structure and magnetic behavior is further evidenced by comparing the lattice parameters at 4~K before and after dehydration for MnFe and CoFe (Table~\ref{table5}). The magnetic structure of dehydrated MnFe is discussed in the main text.

\begin{table*}
\centering
\small
\begin{tabular}{lcccccc}
\toprule
\multirow{2}{*}{Property}
& \multicolumn{2}{c}{MnFe}
& \multicolumn{2}{c}{NiFe}
& \multicolumn{2}{c}{CoFe} \\
\cmidrule(lr){2-3} \cmidrule(lr){4-5} \cmidrule(lr){6-7}
 & Hydrated & Dehydrated & Hydrated & Dehydrated & Hydrated & Dehydrated \\
\midrule
a (\AA) 4K & 10.42(1) & 10.38(1)  & 10.187(4) & 10.19(1) & 10.27(1) & 9.99(1) \\
a (\AA) 300 K & 10.49(1) & 10.44(1) & 10.22(1) & 10.22(1) & 10.30(1) &  10.01(1) \\
$T_C$ (K) & 12 &  9.8 &  21 & 20 & 14  & $\sim$ 4  \\
$\Theta$ (K) & -10 & -12 & 26 & 21  &  -20 & -30 \\
$\mu_{\mathrm{eff}}$ ($\mu_B$)/f.u. & 10.41  & 9.93  & 5.90  & 5.62 & 9.04 &  7.83 \\
$M_s$ ($\mu_B$/f.u.) (5 T) &  9.91 & 7.16  & 5.37   & - & 4.62  & - \\
\bottomrule
\end{tabular}
\caption{Comparison of magnetic properties of selected PBAs before and after dehydration. Magnetic susceptibility was measured under 500 Oe, and Curie–Weiss fitting in the range 150–300 K.}
\label{table5}
\end{table*}

\newpage
\section{Minimal Spin Hamiltonian for MnFe}
\label{sec-hamiltonian}

This section provides the complete derivation of the minimal spin Hamiltonian used in the main text to rationalize the field-induced spin reorientation of MnFe, together with the numerical estimate of the exchange parameters.

\subsection{Hamiltonian without applied magnetic field}

We consider a face-centered cubic (FCC) unit cell containing four Mn atoms and four Fe atoms. The magnetic Hamiltonian includes nearest-neighbor Mn--Fe interactions and second-neighbor Mn--Mn and Fe--Fe interactions:

\begin{equation}
\begin{split}
\mathcal{H} & =
J_{\mathrm{Mn-Fe}} \sum_{\langle i,j \rangle}^{12} \mathbf{S}^{\mathrm{Mn}}_i \cdot \mathbf{S}^{\mathrm{Fe}}_j + J_{\mathrm{Mn-Mn}} \sum_{\langle i,j \rangle}^{6} \mathbf{S}^{\mathrm{Mn}}_i \cdot \mathbf{S}^{\mathrm{Mn}}_j \\
& + J_{\mathrm{Fe-Fe}} \sum_{\langle i,j \rangle}^{6} \mathbf{S}^{\mathrm{Fe}}_i \cdot \mathbf{S}^{\mathrm{Fe}}_j
\end{split}
\end{equation}

\textbf{Fe atoms}
\begin{align*}
\mathbf{S}^{\mathrm{Fe}}_1 &: (0,0,0), \\
\mathbf{S}^{\mathrm{Fe}}_2 &: (0,\tfrac{1}{2},\tfrac{1}{2}), \\
\mathbf{S}^{\mathrm{Fe}}_3 &: (\tfrac{1}{2},0,\tfrac{1}{2}), \\
\mathbf{S}^{\mathrm{Fe}}_4 &: (\tfrac{1}{2},\tfrac{1}{2},0),
\end{align*}

\textbf{Mn atoms}
\begin{align*}
\mathbf{S}^{\mathrm{Mn}}_1 &: (\tfrac{1}{2},0,0), \\
\mathbf{S}^{\mathrm{Mn}}_2 &: (\tfrac{1}{2},\tfrac{1}{2},\tfrac{1}{2}), \\
\mathbf{S}^{\mathrm{Mn}}_3 &: (0,0,\tfrac{1}{2}), \\
\mathbf{S}^{\mathrm{Mn}}_4 &: (0,\tfrac{1}{2},0).
\end{align*}

\subsubsection{Total Magnetic Energy of the Unit Cell}

The total energy of the unit cell is written as

\begin{equation}
E_{\mathrm{tot}} = E_{\mathrm{Mn-Fe}} + E_{\mathrm{Mn-Mn}} + E_{\mathrm{Fe-Fe}}
\end{equation}

\paragraph{Mn--Fe contribution.}
There are 12 distinct Mn--Fe bonds inside the unit cell:

\begin{align}
E_{\mathrm{Mn-Fe}} =
J_{\mathrm{Mn-Fe}} (&
\mathbf{S}^{\mathrm{Fe}}_1\!\cdot\!\mathbf{S}^{\mathrm{Mn}}_1 +
\mathbf{S}^{\mathrm{Fe}}_1\!\cdot\!\mathbf{S}^{\mathrm{Mn}}_3 +
\mathbf{S}^{\mathrm{Fe}}_1\!\cdot\!\mathbf{S}^{\mathrm{Mn}}_4 \nonumber \\
&+
\mathbf{S}^{\mathrm{Fe}}_2\!\cdot\!\mathbf{S}^{\mathrm{Mn}}_2 +
\mathbf{S}^{\mathrm{Fe}}_2\!\cdot\!\mathbf{S}^{\mathrm{Mn}}_3 +
\mathbf{S}^{\mathrm{Fe}}_2\!\cdot\!\mathbf{S}^{\mathrm{Mn}}_4 \nonumber \\
&+
\mathbf{S}^{\mathrm{Fe}}_3\!\cdot\!\mathbf{S}^{\mathrm{Mn}}_1 +
\mathbf{S}^{\mathrm{Fe}}_3\!\cdot\!\mathbf{S}^{\mathrm{Mn}}_2 +
\mathbf{S}^{\mathrm{Fe}}_3\!\cdot\!\mathbf{S}^{\mathrm{Mn}}_3 \nonumber \\
&+
\mathbf{S}^{\mathrm{Fe}}_4\!\cdot\!\mathbf{S}^{\mathrm{Mn}}_1 +
\mathbf{S}^{\mathrm{Fe}}_4\!\cdot\!\mathbf{S}^{\mathrm{Mn}}_2 +
\mathbf{S}^{\mathrm{Fe}}_4\!\cdot\!\mathbf{S}^{\mathrm{Mn}}_4 ).
\end{align}

\paragraph{Mn--Mn contribution.}
There are 6 distinct Mn--Mn second-neighbor pairs:

\begin{align}
E_{\mathrm{Mn-Mn}} =
J_{\mathrm{Mn-Mn}} (&
\mathbf{S}^{\mathrm{Mn}}_1\!\cdot\!\mathbf{S}^{\mathrm{Mn}}_2 +
\mathbf{S}^{\mathrm{Mn}}_1\!\cdot\!\mathbf{S}^{\mathrm{Mn}}_3 +
\mathbf{S}^{\mathrm{Mn}}_1\!\cdot\!\mathbf{S}^{\mathrm{Mn}}_4 \nonumber \\
&+
\mathbf{S}^{\mathrm{Mn}}_2\!\cdot\!\mathbf{S}^{\mathrm{Mn}}_3 +
\mathbf{S}^{\mathrm{Mn}}_2\!\cdot\!\mathbf{S}^{\mathrm{Mn}}_4 +
\mathbf{S}^{\mathrm{Mn}}_3\!\cdot\!\mathbf{S}^{\mathrm{Mn}}_4 ).
\end{align}

\paragraph{Fe--Fe contribution.}
Similarly,

\begin{align}
E_{\mathrm{Fe-Fe}} =
J_{\mathrm{Fe-Fe}} (&
\mathbf{S}^{\mathrm{Fe}}_1\!\cdot\!\mathbf{S}^{\mathrm{Fe}}_2 +
\mathbf{S}^{\mathrm{Fe}}_1\!\cdot\!\mathbf{S}^{\mathrm{Fe}}_3 +
\mathbf{S}^{\mathrm{Fe}}_1\!\cdot\!\mathbf{S}^{\mathrm{Fe}}_4 \nonumber \\
&+
\mathbf{S}^{\mathrm{Fe}}_2\!\cdot\!\mathbf{S}^{\mathrm{Fe}}_3 +
\mathbf{S}^{\mathrm{Fe}}_2\!\cdot\!\mathbf{S}^{\mathrm{Fe}}_4 +
\mathbf{S}^{\mathrm{Fe}}_3\!\cdot\!\mathbf{S}^{\mathrm{Fe}}_4 ).
\end{align}

\subsubsection{Configuration I: Partially Frustrated Collinear State}

We consider the following collinear configuration:

\begin{align*}
\mathbf{S}^{\mathrm{Mn}}_1 &= +S^{\mathrm{Mn}} \hat{a}, \\
\mathbf{S}^{\mathrm{Mn}}_2 &= +S^{\mathrm{Mn}} \hat{a}, \\
\mathbf{S}^{\mathrm{Mn}}_3 &= -S^{\mathrm{Mn}} \hat{a}, \\
\mathbf{S}^{\mathrm{Mn}}_4 &= +S^{\mathrm{Mn}} \hat{a},
\end{align*}

and Fe moments locally antiparallel:

\begin{align*}
\mathbf{S}^{\mathrm{Fe}}_1 &= -S^{\mathrm{Fe}} \hat{a}, \\
\mathbf{S}^{\mathrm{Fe}}_2 &= -S^{\mathrm{Fe}} \hat{a}, \\
\mathbf{S}^{\mathrm{Fe}}_3 &= +S^{\mathrm{Fe}} \hat{a}, \\
\mathbf{S}^{\mathrm{Fe}}_4 &= -S^{\mathrm{Fe}} \hat{a}.
\end{align*}

Since all spins are collinear, scalar products reduce to $\mathbf{S}_i \cdot \mathbf{S}_j = \pm S_i S_j$.

\paragraph{Mn--Fe contribution.}
Counting satisfied and frustrated bonds among the 12 Mn--Fe pairs: 10 bonds are antiferromagnetically satisfied, 2 bonds are frustrated. Therefore,

\begin{equation}
\begin{split}
E_{\mathrm{Mn-Fe}} & = J_{\mathrm{Mn-Fe}} (-10 S^{\mathrm{Mn}} S^{\mathrm{Fe}} +2 S^{\mathrm{Mn}} S^{\mathrm{Fe}}) \\
& = -8 J_{\mathrm{Mn-Fe}} S^{\mathrm{Mn}} S^{\mathrm{Fe}}.
\end{split}
\end{equation}

\paragraph{Mn--Mn contribution.}
Among the 6 Mn--Mn pairs, 3 pairs are parallel and 3 pairs are antiparallel. Thus,

\begin{equation}
E_{\mathrm{Mn-Mn}} = J_{\mathrm{Mn-Mn}} (3 S^2 - 3 S^2) = 0.
\end{equation}

\paragraph{Fe--Fe contribution.}
Among the 6 Fe--Fe pairs, 3 pairs are parallel and 3 pairs are antiparallel. Thus,

\begin{equation}
E_{\mathrm{Fe-Fe}} = J_{\mathrm{Fe-Fe}} (3 S^2 - 3 S^2) = 0.
\end{equation}

\paragraph{Total energy of Configuration I.}
\begin{equation}
\boxed{ E_I = -8 J_{\mathrm{Mn-Fe}} S^{\mathrm{Mn}} S^{\mathrm{Fe}} }
\end{equation}

\subsubsection{Configuration II: Uniform Collinear State}
All Mn moments are aligned along $+\hat{a}$ and all Fe moments along $-\hat{a}$:

\begin{equation}
\boxed{
\begin{split}
E_{II} & = -12 J_{\mathrm{Mn-Fe}} S^{\mathrm{Mn}} S^{\mathrm{Fe}} + 6 J_{\mathrm{Mn-Mn}} (S^{\mathrm{Mn}})^2 \\
&+ 6 J_{\mathrm{Fe-Fe}} (S^{\mathrm{Fe}})^2
\end{split}
}
\end{equation}

\subsubsection{Energy Difference and Stability Condition}
\label{sec-stability}
The energy difference reads

\begin{equation}
E_{II} - E_I = -4 J_{\mathrm{Mn-Fe}} S^{\mathrm{Mn}} S^{\mathrm{Fe}} + 6 J_{\mathrm{Mn-Mn}} (S^{\mathrm{Mn}})^2 + 6 J_{\mathrm{Fe-Fe}} (S^{\mathrm{Fe}})^2.
\end{equation}

Configuration I is energetically favorable when

\begin{equation}
6 J_{\mathrm{Mn-Mn}} (S^{\mathrm{Mn}})^2 + 6 J_{\mathrm{Fe-Fe}} (S^{\mathrm{Fe}})^2 > 4 J_{\mathrm{Mn-Fe}} S^{\mathrm{Mn}} S^{\mathrm{Fe}}.
\end{equation}

Since $J_{\mathrm{Mn-Fe}} > 0$ (antiferromagnetic Mn--Fe coupling), this condition requires $J_{\mathrm{Mn-Mn}}$ (and/or $J_{\mathrm{Fe-Fe}}$) to be positive as well, i.e.\ likewise antiferromagnetic in this convention, and large enough to outweigh the Mn--Fe term. This is consistent with the partially frustrated $\mathbf{k}=(1,0,0)$ ground state observed experimentally at zero field for MnFe, which reflects the geometric frustration of the antiferromagnetically coupled Mn face-centered-cubic sublattice.

\subsection{Hamiltonian with applied magnetic field}

We now include an external magnetic field $\mathbf{B}$ coupling to both Mn and Fe magnetic moments. The Hamiltonian becomes

\begin{equation}
\begin{split}
\mathcal{H} & = J_{\mathrm{Mn-Fe}} \sum_{\langle i,j \rangle}^{12} \mathbf{S}^{\mathrm{Mn}}_i \cdot  \mathbf{S}^{\mathrm{Fe}}_j + J_{\mathrm{Mn-Mn}} \sum_{\langle i,j \rangle}^{6} \mathbf{S}^{\mathrm{Mn}}_i \cdot  \mathbf{S}^{\mathrm{Mn}}_j \\
& + J_{\mathrm{Fe-Fe}} \sum_{\langle i,j \rangle}^{6} \mathbf{S}^{\mathrm{Fe}}_i \cdot \mathbf{S}^{\mathrm{Fe}}_j -
\mathbf{B} \cdot \left( \sum_{i=1}^{4} \mathbf{S}^{\mathrm{Mn}}_i + \sum_{i=1}^{4} \mathbf{S}^{\mathrm{Fe}}_i \right)
\end{split}
\end{equation}

The total energy of the unit cell is therefore

\begin{equation}
E_{\mathrm{tot}} = E_{\mathrm{exchange}} + E_{\mathrm{Zeeman}},
\end{equation}
with the Zeeman contribution

\begin{equation}
E_{\mathrm{Zeeman}} = - \mathbf{B} \cdot \left( \sum_{i=1}^{4} \mathbf{S}^{\mathrm{Mn}}_i + \sum_{i=1}^{4} \mathbf{S}^{\mathrm{Fe}}_i \right).
\end{equation}

We assume a collinear magnetic field along $\hat{a}$: $\mathbf{B} = B \hat{a}$. Since all spins are collinear along $\hat{a}$, scalar products reduce to $\pm S_i S_j$ and $\mathbf{B} \cdot \mathbf{S}_i = \pm B S$.

\subsubsection{Configuration I: Partially Frustrated Collinear State}

\begin{align*}
\mathbf{S}^{\mathrm{Mn}}_1 &= +S^{\mathrm{Mn}} \hat{a}, \quad
\mathbf{S}^{\mathrm{Mn}}_2 = +S^{\mathrm{Mn}} \hat{a}, \quad
\mathbf{S}^{\mathrm{Mn}}_3 = -S^{\mathrm{Mn}} \hat{a}, \quad
\mathbf{S}^{\mathrm{Mn}}_4 = +S^{\mathrm{Mn}} \hat{a},
\end{align*}
\begin{align*}
\mathbf{S}^{\mathrm{Fe}}_1 &= -S^{\mathrm{Fe}} \hat{a}, \quad
\mathbf{S}^{\mathrm{Fe}}_2 = -S^{\mathrm{Fe}} \hat{a}, \quad
\mathbf{S}^{\mathrm{Fe}}_3 = +S^{\mathrm{Fe}} \hat{a}, \quad
\mathbf{S}^{\mathrm{Fe}}_4 = -S^{\mathrm{Fe}} \hat{a}.
\end{align*}

\paragraph{Exchange energy.}
From the previous calculation:
\begin{equation}
E_{\mathrm{exchange}}^{(I)} = -8 J_{\mathrm{Mn-Fe}} S^{\mathrm{Mn}} S^{\mathrm{Fe}}
\end{equation}
(Mn--Mn and Fe--Fe contributions cancel.)

\paragraph{Zeeman energy.}
Total Mn magnetization: $M_{\mathrm{Mn}} = ( +1 +1 -1 +1 ) S^{\mathrm{Mn}} = 2 S^{\mathrm{Mn}}$. Total Fe magnetization: $M_{\mathrm{Fe}} = ( -1 -1 +1 -1 ) S^{\mathrm{Fe}} = -2 S^{\mathrm{Fe}}$. Thus
\begin{equation}
E_{\mathrm{Zeeman}}^{(I)} = - B ( 2 S^{\mathrm{Mn}} - 2 S^{\mathrm{Fe}} ).
\end{equation}

\paragraph{Total energy.}
\begin{equation}
\boxed{ E_{I}(B) = -8 J_{\mathrm{Mn-Fe}} S^{\mathrm{Mn}} S^{\mathrm{Fe}} - 2 B ( S^{\mathrm{Mn}} - S^{\mathrm{Fe}} ) }
\end{equation}

\subsubsection{Configuration II: Uniform Collinear State}

All Mn spins are along $+\hat{a}$ and all Fe spins along $-\hat{a}$.

\paragraph{Exchange energy.}
\begin{equation}
\begin{split}
E_{\mathrm{exchange}}^{(II)} & = -12 J_{\mathrm{Mn-Fe}} S^{\mathrm{Mn}} S^{\mathrm{Fe}} \\
& + 6 J_{\mathrm{Mn-Mn}} (S^{\mathrm{Mn}})^2 + 6 J_{\mathrm{Fe-Fe}} (S^{\mathrm{Fe}})^2
\end{split}
\end{equation}

\paragraph{Zeeman energy.}
Total Mn magnetization: $M_{\mathrm{Mn}} = 4 S^{\mathrm{Mn}}$. Total Fe magnetization: $M_{\mathrm{Fe}} = -4 S^{\mathrm{Fe}}$. Thus
\begin{equation}
E_{\mathrm{Zeeman}}^{(II)} = - B (4 S^{\mathrm{Mn}} - 4 S^{\mathrm{Fe}}).
\end{equation}

\paragraph{Total energy.}
\begin{equation}
\boxed{
\begin{split}
E_{II}(B) & = -12 J_{\mathrm{Mn-Fe}} S^{\mathrm{Mn}} S^{\mathrm{Fe}} + 6 J_{\mathrm{Mn-Mn}} (S^{\mathrm{Mn}})^2 \\
& + 6 J_{\mathrm{Fe-Fe}} (S^{\mathrm{Fe}})^2 - 4 B ( S^{\mathrm{Mn}} - S^{\mathrm{Fe}} )
\end{split}
}
\end{equation}

\subsubsection{Energy Difference}

\begin{equation}
\begin{split}
E_{II}(B) - E_{I}(B) & = -4 J_{\mathrm{Mn-Fe}} S^{\mathrm{Mn}} S^{\mathrm{Fe}} + 6 J_{\mathrm{Mn-Mn}} (S^{\mathrm{Mn}})^2 \\
& + 6 J_{\mathrm{Fe-Fe}} (S^{\mathrm{Fe}})^2 - 2 B ( S^{\mathrm{Mn}} - S^{\mathrm{Fe}} )
\end{split}
\end{equation}

\bibliographystyle{apsrev}
\bibliography{PRB_Draft_1}